\documentclass[sigconf]{acmart}

\usepackage{booktabs}
\usepackage{multirow}
\usepackage{xcolor}
\usepackage{tikz}
\usetikzlibrary{positioning,arrows.meta,fit,calc}
\usepackage{pgfplots}
\pgfplotsset{compat=1.17}
\usepackage{todonotes}
\usepackage{placeins}
\usepackage{adjustbox}
\usepackage{listings}
\newcommand{\method}{\textbf{ExecJudges}}
\newcommand{\methodL}{\method{}-L}

\setcopyright{none}
\renewcommand\footnotetextcopyrightpermission[1]{}
\acmYear{2027}

\begin{document}
\emergencystretch=1.5em

\title{Can Programs Replace LLM Judges? Executable Program Judges for Scalable and Efficient Relevance Labeling}

\author{Eugene Agichtein}
\affiliation{\institution{Emory University}\city{Atlanta}\state{GA}\country{USA}}
\email{eugene.agichtein@emory.edu}
\author{Charles L. A. Clarke}
\affiliation{\institution{University of Waterloo}\city{Waterloo}\state{ON}\country{Canada}}
\email{claclark@gmail.com}
\author{Kaustubh Dhole}\authornote{Work partially done at Emory University.}
\affiliation{\institution{Stanford University}\city{Stanford}\state{CA}\country{USA}}
\email{kdhole@stanford.edu}
\author{William Stern}
\affiliation{\institution{Emory University}\city{Atlanta}\state{GA}\country{USA}}
\email{william.stern@emory.edu}

\begin{abstract}
Large language model (LLM) judges have become a common way to create large-scale relevance judgments for retrieval evaluation.
However, each judgment requires a potentially expensive call to a frontier model, and the judgments are hard to interpret and may not be reproducible. In this paper, we ask whether an LLM can write a readable program for each topic that replaces its per-document judgments. To address these issues, we propose ExecJudges, a method for generating fast, inspectable and scalable judges for automatic relevance labeling.
Specifically, ExecJudges uses a frontier LLM to turn assessment guidelines into executable programs, which rely only on lexical matching, standard NLP libraries like parsers, and pool statistics. An ExecJudges-L variant also uses examples of human labels if available. We evaluate ExecJudges on seven TREC collections by comparing the produced labels with trained NIST assessors, frontier LLM judges, and recently proposed per-topic classifiers. Our results show that ExecJudges agrees with NIST assessors at the level of the Claude Sonnet 5.5 LLM judges on TREC DL 2019--2021. With a small number of human labels per topic, the ExecJudges-L variant agrees with NIST more than per-topic classifiers with the same labels on all seven collections, and either exceeds the agreement of LLM judges provided the same labels or is within about 0.05 of it in Krippendorff's $\alpha$. The labels rank the official TREC DL runs with Kendall's $\tau$ of 0.87 to 0.94, and writing the programs incurs a one-time cost of \$0.17 to \$0.49 per topic, with no additional costs for grading. Our results suggest that ExecJudges offers a reproducible, inspectable, efficient and cost-effective alternative to LLM judges for labeling deep result pools for system evaluation or ranking improvements.

\end{abstract}


\maketitle
\pagestyle{plain}

\section{Introduction}

Relevance judgments are essential for evaluating information retrieval systems. Obtaining them from human assessors is expensive and difficult to scale. Large language models (LLMs) have emerged as a promising alternative. Their relevance labels often approach the agreement between human assessors~\cite{thomas2024large,upadhyay2025large}. However, relying on LLMs as relevance judges introduces several limitations. First, an expensive and often GPU-dependent model must evaluate each document. Assessment costs therefore grow with the size of the document pool and with every new retrieval run. Second, the decision process of an LLM judge is largely opaque. Even when the LLM provides an explanation, it is difficult to verify which criteria determined its judgment. Third, irrelevant or adversarial content can influence LLM judges. Query terms and instructions embedded in an assessed document are examples of such content~\cite{alaofi2024fooled}. These limitations raise concerns about the reliability, reproducibility, and suitability of LLM-generated relevance judgments as substitutes for human assessments~\cite{faggioli2023perspectives,soboroff2025dont}.

Relevance assessment involves repeatedly applying the same topic-specific criteria to different documents. For a given topic, assessors must determine whether a document addresses the requested subject, provides an answer, and satisfies the conditions for higher relevance grades. Although these criteria vary across topics, their interpretation and application need not be reconstructed independently for every document. Recent work on executable rubrics converts the natural-language evaluation criteria of LLM judges into explicit programs~\cite{dhole2026execrubrics}. These programs replace repeated LLM judgments with deterministic scoring procedures that people can inspect. Motivated by this direction, we ask a related question for information retrieval:~\textbf{\textit{Can an LLM write a readable, generalizable relevance program for each topic that replaces its per-document relevance judgments?}} Such an approach could keep the topic-specific reasoning of LLMs. It would also reduce inference costs and make the assessment logic explicit and open to inspection.

We propose \method{}, a method that uses LLMs to generate topic-specific relevance assessors in the form of human-readable Python programs. Each program uses deterministic operations, including lexical matching, text parsing, and pool-level statistics, to evaluate the evidence needed for a relevance judgment. Two design choices distinguish our approach. First, the programs answer the conditions of the official assessment guidelines. These conditions include whether a passage is on topic, contains an answer, provides an exact answer, or is dedicated to the query. A fixed grading function then maps the answers to relevance grades. This function follows the structure of the assessment scale. Second, we generate several programs for each topic independently and average their grades. Averaging reduces the effect of errors in single programs and reduces over-grading. In this way, \method{} uses the LLM once per topic to construct a reusable assessment procedure. People can run, inspect and revise this procedure. Specifically, our contributions are:
\begin{itemize}
\item We propose \method{}, a new method that uses an LLM to generate relevance-grading programs. The programs label deep document pools without further LLM calls. \method{} agrees with NIST at the level of the Sonnet judges on TREC DL 2019--2021 (Sections~\ref{sec:method} and~\ref{sec:results}) at a one-time cost of \$0.17 to \$0.49 per topic.
\item We investigate adding a few human labels of each topic to the writer prompt. With 16 labels, \methodL{} agrees with NIST more than per-topic classifiers on all seven collections, and more than frontier LLM judges given the same labels on TREC DL 2019--2022 (Sections~\ref{sec:method} and~\ref{sec:results}).
\item We show that \method{} ranks the systems similarly to the official ranking based on the NIST relevance labels with Kendall's $\tau$ of 0.87 to 0.94 (Sections~\ref{sec:setup} and~\ref{sec:results}).
\item We analyze the number of programs, per-topic and collection-level programs, rubric induction, and the choice of labeled examples (Section~\ref{sec:analysis}). Appendix~\ref{app:release} reports all prompts and labeling details.
\end{itemize}

\section{Related Work}
\label{sec:related}

We now review related work to contextualize our contributions.

\textit{\textbf{LLM relevance judges:}} The past few years have seen an increasing use of large language models (LLMs) as automated relevance assessors. Faggioli et al.~\cite{faggioli2023perspectives} provided an early framework for LLM-based relevance judgment. Their framework outlines a spectrum of human-LLM collaboration from human-assisted assessment to fully automated evaluation. \citet{thomas2024large} showed that LLMs can predict searcher preferences. Their LLM relevance judgments were comparable to those of human assessors at a much lower annotation cost. Subsequently, Upadhyay et al.~\cite{upadhyay2025large} introduced UMBRELA, an open-source relevance assessment framework based on GPT-4o. UMBRELA reproduced the Bing relevance assessment method and showed promising results on the TREC Deep Learning benchmarks. Others extended automated assessment to retrieval-augmented generation (RAG)~\cite{thakur2025support,dhole2025conqret} by comparing LLMs with human judges. \method{} uses the same guidelines as these judges and removes the per-document model call. 

\textit{\textbf{Per-topic relevance assessment:}}
Prior work has explored topic-specific supervision and artifacts to improve relevance assessment. Gienapp et al.~\cite{gienapp2026topic} train lightweight classifiers on 64--256 human judgments per topic. These classifiers agree with human assessors more than prompted LLMs. They need topic-specific annotations. Otero et al.~\cite{otero2026hybrid} generate a reusable relevance description for each topic. An LLM then applies this description to judge documents. Farzi et al.~\cite{farzi2024pencils} use per-query exam questions to grade passages. RubricRAG~\cite{dhole2026rubricrag} retrieves related rubrics to generate query-specific evaluation criteria. Qworld~\cite{su2026qworld} recursively generates fine-grained, question-specific evaluation criteria for LLM responses. Wang et al.~\cite{wang2023boolean} generate topic-specific Boolean queries that provide high-precision but low-recall filtering. These approaches prepare labels, examples or other artifacts for each topic. Most of them still apply a model to every document.

\textit{\textbf{Limitations of LLMs as relevance assessors:}}
LLM relevance judges also have known limitations. Soboroff~\cite{soboroff2025dont} argued against relying on LLM-generated judgments for TREC-style evaluations. He emphasized the importance of independent human assessments. Similarly, Clarke and Dietz~\cite{clarke2024llm} highlighted fundamental limitations such as susceptibility to evaluation gaming, circularity, and distorted system rankings. Two practical problems have received less attention. LLM relevance assessments are hard to interpret. Their computational cost makes large-scale assessment slow and expensive.

\textit{\textbf{Symbolic distillation and program synthesis:}} A parallel line of work distills the behavior of black-box models into explicit symbolic rules. Examples include circuit discovery~\cite{rai2026data}, symbolic models distilled from trained networks~\cite{cranmer2020discovering}, and the translation of predictions into rules or reasoning trails~\cite{armgaan2024graphtrail,rorseth2026ruben}. Most of these methods fit rules to a trained model or to its predictions. More recent methods have an LLM write programs directly. ALCHEmist~\cite{huang2024alchemist} has an LLM write labeling programs for classification tasks to label data without model calls. Most recently, \citet{dhole2026execrubrics} translated natural-language rubrics into executable Python rubrics to give faster and verifiable preference annotations of long-form responses. Unlike previous works, \method{} uses an LLM to convert general assessment guidelines into labeling programs, without requiring explicit per-topic rubrics or LLM relevance labels.

\vspace{-6pt}
\section{The \method{} Method}
\label{sec:method}

\method{} replaces the per-document calls of an LLM judge with programs that an LLM writes once for each topic.

\subsection{Writer inputs and tools}

The program writer is an LLM that is called once per topic. It receives the topic, the grading guidelines and a few examples, and returns programs that grade documents. The programs then label every document of the pool without further model calls. Figure~\ref{fig:pipeline} contrasts this process, on the right, with LLM judges, on the left. Our implementation uses Claude Opus 5.5 as the program writer. For each topic, the program writer receives the following inputs:

\begin{figure*}[t]
\centering
\resizebox{\textwidth}{!}{\begin{tikzpicture}[
  font=\scriptsize,
  panel/.style={draw, rounded corners=4pt, inner sep=4pt},
  box/.style={draw, rounded corners=2pt, align=left, inner sep=3pt, fill=white, anchor=north west},
  call/.style={draw, rounded corners=2pt, align=center, inner sep=2pt, fill=red!8, anchor=north west, text width=1.75cm},
  arr/.style={-{Stealth[length=1.5mm]}, semithick}]
\node[box, text width=2.1cm] (ta) at (0,0) {\textbf{Topic}\\query, grade definitions, labeled examples};
\node[box, text width=2.1cm] (pa) at (0,-1.45) {\textbf{Document pool}\\$d_1, d_2, \ldots, d_N$};
\node[call] (c1) at (3.0,0) {LLM judge call\\topic + $d_1$};
\node[call] (c2) at (3.0,-0.8) {LLM judge call\\topic + $d_2$};
\node at (3.95,-1.75) {$\vdots$};
\node[call] (cn) at (3.0,-2.05) {LLM judge call\\topic + $d_N$};
\node[box] (g1) at (5.45,-0.08) {grade};
\node[box] (g2) at (5.45,-0.88) {grade};
\node[box] (gn) at (5.45,-2.13) {grade};
\foreach \c/\g in {c1/g1, c2/g2, cn/gn} \draw[arr] (\c.east) -- (\g.west);
\coordinate (busa) at (2.65,0);
\draw[semithick] (ta.east) -- (ta.east -| busa);
\draw[semithick] (pa.east) -- (pa.east -| busa);
\draw[semithick] (busa |- c1.west) -- (busa |- cn.west);
\foreach \c in {c1, c2, cn} \draw[arr] (busa |- \c.west) -- (\c.west);
\node[align=left, anchor=north west, text width=6.2cm] (na) at (0,-3.0) {%
$\bullet$ One LLM call per document: the cost grows with the pool.\\
$\bullet$ The reasoning behind a grade stays inside the model.};
\node[panel, fit=(ta) (pa) (c1) (cn) (g1) (gn) (na), label={[font=\footnotesize\bfseries]above:(a) Relevance assessment with an LLM judge}] {};
\node[box, text width=2.1cm] (tb) at (7.25,0) {\textbf{Topic}\\query, grade definitions, labeled examples};
\node[call, fill=blue!10, text width=1.9cm] (w) at (10.0,-0.05) {LLM program writer\\for each topic};
\node[box, fill=blue!6, text width=5.0cm] (pr) at (12.55,0) {\textbf{$k$ programs} (we use $k=3$), written independently from the same prompt. Each program returns four answers per document: \texttt{on\_subject}, \texttt{some\_answer}, \texttt{exact\_answer}, \texttt{dedicated}.};
\draw[arr] (tb.east |- w.west) -- (w.west);
\draw[arr] (w.east) -- (w.east -| pr.west);
\node[box, text width=2.1cm] (pb) at (7.25,-1.45) {\textbf{Document pool}\\$d_1, d_2, \ldots, d_N$};
\node[box, fill=gray!6, text width=3.6cm] (run) at (10.0,-1.45) {\textbf{Run the programs on a CPU, without model calls.} Grade per program: 0 if not on subject; 1 if no answer; 3 if exact and dedicated; 2 otherwise.};
\node[box, fill=green!6, text width=3.65cm] (lab) at (14.05,-1.45) {\textbf{Label} = floor of mean of the grades\\[1pt]
\begin{tabular}{@{}l@{\ \ }c@{\ \ }l@{}}
$d_1$ & 3, 2, 3 & $\rightarrow \lfloor 2.67\rfloor = 2$\\
$d_2$ & 0, 0, 1 & $\rightarrow \lfloor 0.33\rfloor = 0$\\
$d_3$ & 1, 1, 2 & $\rightarrow \lfloor 1.33\rfloor = 1$
\end{tabular}};
\draw[arr] (pb.east |- run.west) -- (run.west);
\draw[arr] ([xshift=0.55cm]pr.south west) -- ([xshift=0.55cm]pr.south west |- run.north);
\draw[arr] (run.east |- lab.west) -- (lab.west);
\node[align=left, anchor=north west, text width=10.4cm] (nb) at (7.25,-3.0) {%
$\bullet$ One writer call per program and topic. The programs then grade any number of documents.\\
$\bullet$ The rules behind each grade are code that people can read.\\
$\bullet$ \methodL{} adds a few labeled passages of the topic to the writer prompt.};
\node[panel, fit=(tb) (w) (pr) (pb) (run) (lab) (nb), label={[font=\footnotesize\bfseries]above:(b) Relevance assessment with \method{}}] {};
\end{tikzpicture}}
\caption{Relevance assessment with an LLM judge (a) and with \method{} (b). The LLM judge is called for every document. \method{} calls the LLM once per topic to write programs, and the programs grade all documents.}
\label{fig:pipeline}
\end{figure*}
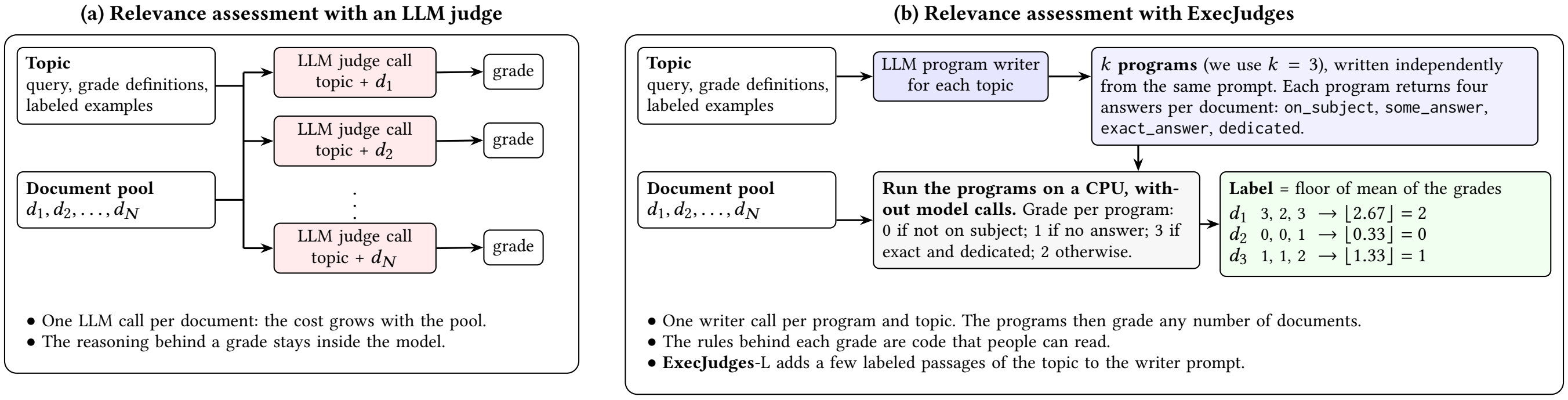

\begin{itemize}
    \item \textbf{Topic information:} The topic as presented to the human assessors, consisting of the query for TREC Deep Learning, and the title, description, and narrative for TREC-COVID and Robust04.

    \item \textbf{Relevance grading guidelines:} The official assessor grade definitions, reproduced verbatim from the corresponding track guidelines.

    \item \textbf{Grade distribution:} A calibration statement that gives the share of each relevance grade. For DL 2019 and 2020, the shares come from the NIST labels of the DL 2019 and 2020 topics. For the other collections, they come from the example topics.

    \item \textbf{Labeled examples:} One NIST-labeled passage per relevance grade from each of a few example topics of the same collection. We use three example topics. DL 2020 uses the DL 2019 example topics. Topics included in the prompt or used for training are excluded from all evaluations.

    \item \textbf{Unlabeled examples:} A few unlabeled passages sampled from the target topic's document pool to illustrate the input text format. We use five passages.

    \item \textbf{Tool-use policy:} A specification of the permitted libraries, operations, and execution constraints for the generated programs.
\end{itemize}

The prompt contains only a few examples. This choice follows the in-context learning literature, in which three to five examples have been shown to be a successful setting for code generation tasks~\cite{austin2021program,li2024acecoder,oskooei2025manyshot}. The grade distribution statement follows the guidance on instructing non-expert assessors with clear calibration and mapping of utility to relevance levels~\cite{sormunen2002liberal,alonso2012crowdsourcing}. Without \methodL{} labels, the program writer never sees a relevance label of the topic for which it writes a program.

The programs use only deterministic, lightweight operations. Grading is therefore cheap and gives the same labels on every run. The programs may call the Python standard library and a small toolkit. The toolkit tokenizes text with Snowball stemming. It computes BM25 scores and percentiles within the pool of the topic~\cite{robertson2009bm25}. It also matches phrases within a window, splits sentences, and parses numbers, dates and units. Two helpers measure how much of a passage supports given terms and how much of it is unrelated text. The programs run in a sandbox with time and memory limits. The sandbox blocks network access and model calls.

\subsection{Program structure and consensus labeling}

Each program answers the questions that the official grade definitions pose about a document. A fixed formula then maps the answers to a grade. For example, for the query \emph{when was the salvation army founded}, a program checks whether a passage is about the Salvation Army, whether it states when the organization was founded, and whether it is mainly about this question. Figure~\ref{fig:example} shows this topic. This design keeps the grading scale outside the generated code. The writer only decides what counts as evidence for each question.

Each program defines one function, \texttt{assess}. This function receives the query and the whole pool of the topic. It returns one record of answers for each document. On the 0--3 scale of TREC Deep Learning, the answers are \texttt{on\_subject}, \texttt{some\_answer}, \texttt{exact\_answer} and \texttt{dedicated}. The formula follows the NIST definitions. A passage that is not on the subject gets grade 0. A passage on the subject without an answer gets grade 1. A passage with an exact answer that is dedicated to the query gets grade 3. Every other passage gets grade 2. On the 0--2 scales of TREC-COVID and Robust04, the answers are \texttt{relevant} and \texttt{key}. These two answers correspond to the definitions of grades 1 and 2 in the track guidelines. The prompt asks the writer to set an answer to true only on explicit evidence. Missing evidence therefore lowers the grade.

We run each new program on test passages from the pool. If the program fails or violates the tool policy, the writer gets one chance to repair it. We allow a single repair attempt, mainly to handle possible API or connection timeouts.

\method{} labels a document with the consensus of several independently written programs. Single programs make different errors and tend to grade too high. Averaging their grades reduces both effects. Specifically, the writer writes $k$ programs independently from the same prompt. Program $i$ assigns the grade $g_i(d)$ to document $d$. The label of the document is the mean grade, rounded down: $\ell(d) = \lfloor \frac{1}{k}\sum_{i=1}^{k} g_i(d) \rfloor$. Our implementation uses $k=3$, because more programs added little agreement in our experiments on the development and pilot collections (Section~\ref{sec:analysis}). We chose the rounding rule after a control experiment on the same collections. Appendix~\ref{app:details} describes this experiment. We fixed $k$ and the rounding rule before any run on the test collections. Figure~\ref{fig:example} shows the main method on one DL 2019 topic.

\begin{figure*}[t]
\centering
\begin{tikzpicture}[node distance=0.25cm and 0.42cm,
  box/.style={draw, rounded corners=2pt, align=left, inner sep=3pt, font=\scriptsize, anchor=north},
  arr/.style={-{Stealth[length=1.8mm]}, thick}]
\node[box, text width=3.9cm] (a) {%
{\bfseries Topic and grade definitions}\\[2pt]
\emph{DL 2019 topic 962179:} \textit{when was the salvation army founded}\\[2pt]
{[3]} dedicated to the query and contains the exact answer\\
{[2]} some answer, possibly unclear or hidden among extraneous information\\
{[1]} related, does not answer\ \ {[0]} irrelevant\\[2pt]
The prompt turns these definitions into four questions per document: \texttt{on\_subject}, \texttt{some\_answer}, \texttt{exact\_answer}, \texttt{dedicated}.};
\node[box, text width=4.6cm, right=of a.north east, anchor=north west] (b) {%
{\bfseries Program 1} (shortened)\\[2pt]
{\ttfamily\tiny
SA = re.compile(r"salvation\textbackslash s+army", re.I)\\
KEY\_YEARS = \{1865, 1878\}\\
def assess(query, results):\ \ \# whole pool\\
\ \ for r in results:\\
\ \ \ \ text = r["text"]\\
\ \ \ \ on\_subject = bool(SA.search(text)) or ...\\
\ \ \ \ for s in jk.sentences(text):\\
\ \ \ \ \ \ if (STRONG\_FOUND.search(s)\\
\ \ \ \ \ \ \ \ \ \ \ \ and jk.years(s) \& KEY\_YEARS):\\
\ \ \ \ \ \ \ \ strong\_sent\_key = True\\
\ \ \ \ some = strong\_sent\_key or found\_sent\_year\\
\ \ \ \ exact = strong\_sent\_key\\
\ \ \ \ dedicated = (jk.unrelated\_share(query, text)\\
\ \ \ \ \ \ \ \ \ \ \ \ \ \ \ \ \ < 0.45 and focus)\\
\ \ \ \ out.append(\{"on\_subject": on\_subject, ...\})}};
\node[box, text width=6.0cm, right=of b.north east, anchor=north west] (c) {%
{\bfseries Answers, grade formula and label}\\[2pt]
$g_k(d)=0$ if not on subject; 1 if no answer; 3 if exact and dedicated; 2 otherwise.\\
$\ell(d)=\left\lfloor \frac{1}{3}\sum_{k=1}^{3} g_k(d)\right\rfloor$\\[3pt]
$d_1$: \textit{The Salvation Army was founded in London's East End in 1865 by one-time Methodist minister William Booth \ldots}\\
program 1 answers: on subject, some, exact, dedicated; grades $(3,3,3)\rightarrow\ell=3$ (NIST 3)\\[2pt]
$d_2$: \textit{\ldots a quasi-military command structure in 1878 when the title, `The Salvation Army', was brought into use.}\\
program 1 answers: on subject only; grades $(1,0,1)\rightarrow\ell=\lfloor 0.67\rfloor=0$ (NIST 0)};
\draw[arr] ($(a.north east)+(0,-0.6)$) -- ($(b.north west)+(0,-0.6)$);
\draw[arr] ($(b.north east)+(0,-0.6)$) -- ($(c.north west)+(0,-0.6)$);
\end{tikzpicture}
\caption{\method{} (Figure~\ref{fig:pipeline}b) on DL 2019 topic 962179, with the grade formula and the rounding of the mean. Passage $d_2$ mentions the Salvation Army without its founding date. Rounding down gives grade 0, the NIST grade.}
\label{fig:example}
\end{figure*}

\subsection{Optional topic labels}

When a few human labels of a topic exist, \methodL{} uses them to show the writer where the assessors drew the lines between the grades. The writer prompt then contains a small set of judged passages of the topic with their NIST grades. The prompt asks the writer to encode general rules that give these passages their grades. The rules must not refer to the passages by their wording. Everything else is the same as in the main method. The programs grade every passage of the pool, including the labeled passages.

We compare two ways to choose the passages to label. The first shows how much labels can help. The second is one that an assessor could apply in practice. \emph{Stratified sampling} takes the same number of passages from each NIST grade. Our implementation uses 1 to 4 passages per grade, for 4 to 16 labels per topic. Our implementation uses 5 passages per grade plus one passage from the largest grade to reach 16 labels on the 0--2 collections. Smaller label sets are subsets of larger ones. Stratified sampling uses the NIST grades. It therefore measures how much a given number of labels can help. \emph{BM25 top $k$} labels the $k$ judged passages with the highest BM25 scores. An assessor who judges the top of a lexical ranking would label these passages first. Our implementation uses $k=3$ for \methodL{}, following the same in-context learning setting of three to five examples, and $k=16$ and $k=64$ for the classifier. The original classifier paper uses 64 labels as its smallest training size. When we use \methodL{} labels as qrels, the labeled passages keep their human grades.

\subsection{Ablation variants}

Two variants test the main design choices of a per-topic program. \method{}-1P tests the value of averaging. It uses only the first program of each topic. \method{}-R tests whether an explicit rubric helps. Its writer first induces 3 to 9 criteria for the topic. Each criterion carries the grade level that requires it. The writer then writes one check function per criterion. The grade is the highest level whose criteria all pass.

Two further variants test whether the programs must be specific to a topic. \method{}-Global is one program for all topics. \method{}-GR is one rubric with one set of check functions for all topics. These two variants label new topics without any LLM call.

\section{Experimental Setup}
\label{sec:setup}

We evaluate \method{} on seven TREC collections against LLM judges and supervised per-topic methods, by label agreement, system-ranking agreement and cost.

\subsection{Collections and protocol}
\label{sec:setup-protocol}

We divide the collections into development, pilot and test collections. This division ensures that the test results do not depend on choices made during method design. DL 2019 and 2020~\cite{craswell2019overview,craswell2020overview} serve for development. DL 2021~\cite{craswell2021overview} serves as the pilot test. The test collections are DL 2022~\cite{craswell2022overview}, DL 2023~\cite{craswell2023overview}, TREC-COVID~\cite{voorhees2020treccovid} and Robust04~\cite{voorhees2004robust}. DL grades range from 0 to 3. TREC-COVID and Robust04 grades range from 0 to 2. We load all collections with ir\_datasets~\cite{macavaney2022irdatasets}. We evaluate every method on the same topics. Table~\ref{tab:collstats} in Appendix~\ref{app:release} gives the size of each collection. The same appendix lists the topics that we excluded and the reasons.

We separated prompt development from testing. We developed the program writer prompts on the development collections and then froze them. DL 2021 served as the pilot test. Appendix~\ref{app:details} lists the changes that we made after the pilot. We ran \method{} and \methodL{} once on each test collection with the frozen prompts. The runs marked $^\dagger$ in Table~\ref{tab:main} were made separately, once each. For DL 2023, we have only these separate runs. API costs use October 2026 prices.

\subsection{Baselines}

We compare \method{} with LLM judges, with supervised per-topic methods, and with baselines that use no LLM. All baselines label the same passages as \method{}. The label-based baselines receive the same labels as \methodL{}.

\begin{itemize}
\item \emph{LLM judges}: We use Claude Sonnet 5.5 and Claude Haiku 5.5 as judges. The judges receive the same inputs as the program writer. The pointwise judge uses the UMBRELA prompt~\cite{upadhyay2025large} with one passage per call. The listwise judge grades a numbered list of up to 100 passages of the pool per request. Table~\ref{tab:main} reports the listwise Sonnet judge on all collections. Section~\ref{sec:setup-protocol} describes which of these runs were made separately.
\item \emph{LLM judges with topic labels}: These are the same judges with the labeled passages of \methodL{} added to the prompt.
\item \emph{Topic classifier}~\cite{gienapp2026topic}: The tables call this baseline Classifier. It trains one LoRA adapter per topic on monoT5-base with the same labels as \methodL{}. With the released training settings, it labels almost every passage relevant. We therefore tuned three settings on 8 DL 2019 topics. This tuning favors the classifier on DL 2019. Appendix~\ref{app:details} gives the details.
\item \emph{One-shot labeling}~\cite{macavaney2023oneshot}: The tables call this baseline One-shot. DuoT5-3B compares each passage with the highest-graded passage of the 16-label set, as in the released autoqrels code. Cut points turn the DuoT5 score into grades. One-shot (1) fits the cut points on the labeled passages of other collections. One-shot (16) fits them on the 16 labels of the topic. One-shot labeling was designed to label the unjudged documents of a pool. Our reuse experiment tests it in that use. On Robust04 and TREC-COVID, we cut each document to its first 230 tokens. This cut keeps both documents within the 512-token input of the released code.
\item \emph{Without an LLM}: BM25 cutoffs rank the pool by BM25 and cut the ranking at the grade proportions of the example topics. Embedding cutoffs apply the same cut to the cosine similarity of bge-small-en-v1.5 embeddings~\cite{xiao2024cpack}. The lexical classifier is a logistic regression on the features of our toolkit. We train it on the Sonnet labels of DL 2019 and 2020.
\end{itemize}

The supervised baselines and \methodL{} use the labels in different ways. The supervised baselines fit a model or cut points to the labels. The program writer uses the labels to adjust rules that it derives from the grading guidelines.

\subsection{Measures}

We measure label agreement with ordinal Krippendorff's $\alpha$~\cite{krippendorff2018content}. Gienapp et al. use binary $\alpha$ at grade $\geq$ 1. We also report this measure. We pool agreement over all evaluation passages of a collection. No method is evaluated on a passage whose label it saw. Our implementation uses the residual passages for evaluation: the judged passages of each topic without its 16 stratified labels and its BM25 top passages, 3 on DL 2019--2021 and 16 on the test collections.

We measure how well the labels preserve the ranking of retrieval systems with Kendall's $\tau_b$ and with $\tau_{AP}$~\cite{yilmaz2008tauap}. Stars in the tables mark $p<0.05$ and $p<0.01$. For $\alpha$, the test is a paired t-test over topics of per-topic $\alpha$. For $\tau$, the stars mark topic-bootstrap intervals that exclude 0. Our implementation uses 95\% paired-bootstrap intervals of the difference in pooled $\alpha$ for the intervals in the text. We do not correct for multiple comparisons.

\section{Results}
\label{sec:results}

Programs written once per topic reach the label agreement of frontier LLM judges on TREC DL and exceed supervised per-topic methods when a few labels are available.

\begin{table*}[t]
\caption{Main results: ordinal $\alpha$ against NIST. Sonnet is the listwise judge. Stars: against \method{} (no labels) or \methodL{} (16 labels). $^\dagger$Separate single runs on matched passages, without tests. TREC-COVID: 36 topics with labels. --: not run.}
\label{tab:main}
\small
\setlength{\tabcolsep}{4pt}
\begin{adjustbox}{max width=\textwidth}
\begin{tabular}{@{}lccccccc@{}}
\toprule
Method & DL 2019 & DL 2020 & DL 2021 & DL 2022 & DL 2023 & Robust04 & TREC-COVID \\
\midrule
\multicolumn{8}{@{}l}{\emph{No labels}} \\
\method{} & .613 & .483 & .653 & .549 & .490$^\dagger$ & .392 & .508 \\
Sonnet, listwise & .559$^\dagger$ & .496 & .635 & .542$^\dagger$ & .538$^\dagger$ & .482$^\dagger$ & .679$^\dagger$ \\
One-shot (1) & .489 & .439 & .490$^{**}$ & .461$^{**}$ & -- & $-$.515$^{**}$ & .271$^{**}$ \\
\midrule
\multicolumn{8}{@{}l}{\emph{16 labels}} \\
\methodL{} & .756 & .649 & .750 & .702 & .607$^\dagger$ & .424 & .626 \\
Sonnet, listwise & .626$^\dagger$ & .531$^{**}$ & .713$^{*}$ & .591$^\dagger$ & .604$^\dagger$ & .473$^\dagger$ & .678$^\dagger$ \\
Classifier & .706 & .543$^{*}$ & .665$^{**}$ & .573$^{**}$ & .493 & .131$^{**}$ & .506$^{**}$ \\
One-shot (16) & .577$^{**}$ & .459$^{**}$ & .593$^{**}$ & .532$^{**}$ & -- & $-$.063$^{**}$ & .393$^{**}$ \\
\midrule
\multicolumn{8}{@{}l}{\emph{64 labels}} \\
\methodL{} & .785$^\dagger$ & .724$^\dagger$ & .826$^\dagger$ & .734$^\dagger$ & .651$^\dagger$ & .505$^\dagger$ & .613$^\dagger$ \\
Sonnet, listwise & .709$^\dagger$ & .638$^\dagger$ & .767$^\dagger$ & .647$^\dagger$ & .665$^\dagger$ & .491$^\dagger$ & .686$^\dagger$ \\
\bottomrule
\end{tabular}
\end{adjustbox}
\end{table*}

\subsection{Main results}
\label{sec:main}

Without labels, \method{} agrees with NIST labels at the level of the Sonnet judges on the development and pilot collections. Its differences from the pointwise and listwise Sonnet judges are within the confidence intervals on DL 2019--2021. On DL 2019--2022, \method{} is within 0.054 of the listwise Sonnet judge. It is below this judge on DL 2023, Robust04 and TREC-COVID. Table~\ref{tab:main} shows these results. Tables~\ref{tab:pilot} and~\ref{tab:diffs} in Appendix~\ref{app:more} list the intervals. Without labels, the writer has to infer from the guidelines alone where the assessors drew the lines between the grades. Section~\ref{sec:labels} shows that a few labels of the topic supply this information.

\begin{table}[t]
\caption{Test collections: ordinal / binary $\alpha$ on the documents that the pointwise Sonnet judge labeled. (16): 16 labels per topic. Stars: against Sonnet.}
\label{tab:sonnetsample}
\footnotesize
\setlength{\tabcolsep}{1.8pt}
\begin{adjustbox}{max width=\columnwidth}
\begin{tabular}{@{}lccc@{}}
\toprule
 & DL 2022 & Robust04 & TREC-COVID \\
 & (7,402 docs) & (3,950) & (1,794) \\
\midrule
Sonnet, pointwise & .577 / .434 & .500 / .491 & .693 / .676 \\
\method{} & .513$^{*}$ / .412 & .367$^{**}$ / .364$^{**}$ & .496$^{**}$ / .480$^{**}$ \\
\methodL{} (16) & .628$^{*}$ / .579$^{**}$ & .442 / .436 & .661 / .621$^{*}$ \\
\bottomrule
\end{tabular}
\end{adjustbox}
\end{table}

A significance test on the test collections confirms this order. We compare \method{} with the pointwise Sonnet judge on the documents that the judge labeled. Without labels, \method{} is significantly below the judge on DL 2022, Robust04 and TREC-COVID at $p<0.05$. With labels, \methodL{} is significantly above the judge on DL 2022. Its differences from the judge on Robust04 and TREC-COVID are not significant. Table~\ref{tab:sonnetsample} lists these results, with binary $\alpha$. With labels, the programs are therefore no longer significantly below the judge on any of these collections.

\methodL{} agrees with NIST labels more than an LLM judge with the same labels on TREC DL. On DL 2019--2022, \methodL{} is above the listwise Sonnet judge with the same labels by 0.037 to 0.130. On Robust04 and TREC-COVID, it is below this judge by about 0.05. Table~\ref{tab:main} shows these results, including the values with larger label sets. Both document collections have long documents and topics with narratives. Relevant documents are also rare on Robust04. These conditions appear to be harder for lexical programs than passage ranking.

With the same human labels, \methodL{} agrees with NIST labels more than both supervised baselines on every collection. This result holds for passage and document collections and for both grading scales. All three methods receive the same 16 labels per topic. The difference from the classifier is significant on five of the seven collections. The exceptions are DL 2019 and DL 2023. The DL 2023 value of \methodL{} comes from a separate run without a test. The difference from One-shot is significant on all six collections where we ran One-shot. Table~\ref{tab:main} shows these results. Table~\ref{tab:main-bin} in Appendix~\ref{app:more} gives binary $\alpha$ for the same methods. A likely reason is that 16 labels are too few to fit a supervised model and its cut points. The program writer uses the labels only to adjust rules that it derives from the grading guidelines. The analysis of the cut points in Section~\ref{sec:cls} supports this explanation.

Some cells of Table~\ref{tab:main} come from separate single runs, marked $^\dagger$. In those runs, the programs with labels differ from our runs by up to 0.075. For comparisons that involve these cells, we therefore treat differences below 0.04 as ties. All other comparisons use the significance tests of Section~\ref{sec:setup}.

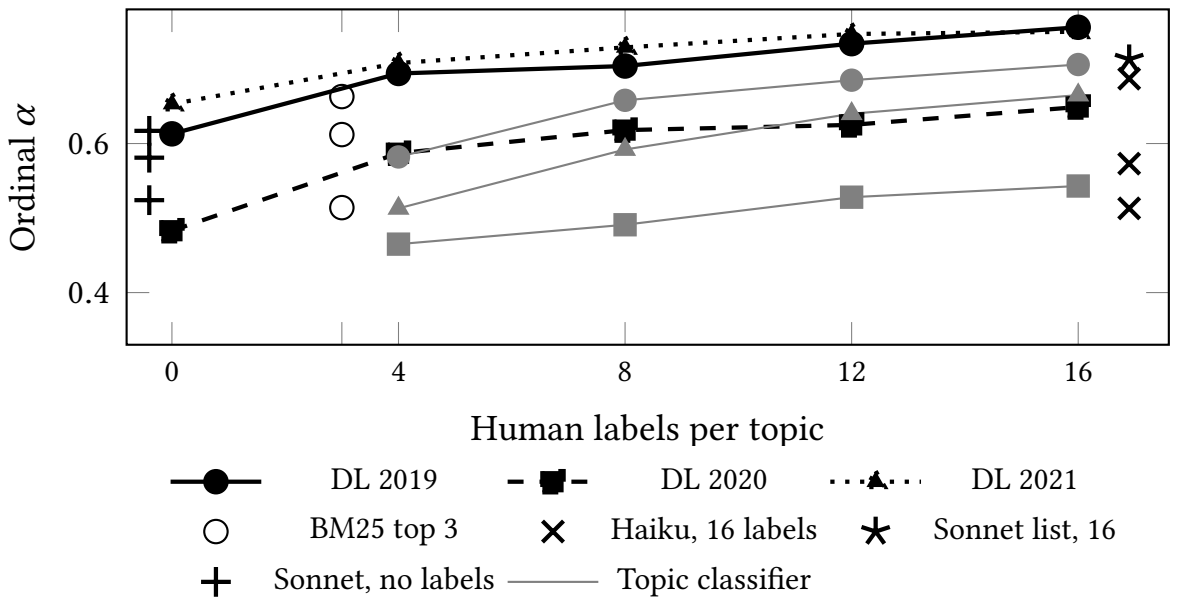
\begin{figure}[t]
\centering
\begin{tikzpicture}
\begin{axis}[width=\columnwidth, height=3.8cm, xlabel={Human labels per topic}, ylabel={Ordinal $\alpha$},
  xtick={0,3,4,8,12,16}, xticklabels={0,,4,8,12,16}, xmin=-0.8, xmax=17.6, ymin=0.33, ymax=0.78,
  legend style={font=\scriptsize, at={(0.5,-0.30)}, anchor=north, legend columns=3, draw=none},
  tick label style={font=\scriptsize}, label style={font=\footnotesize}]
\addplot[mark=*, thick] coordinates {(0,0.613) (4,0.694) (8,0.704) (12,0.734) (16,0.756)}; \addlegendentry{DL 2019}
\addplot[mark=square*, thick, dashed] coordinates {(0,0.483) (4,0.587) (8,0.618) (12,0.625) (16,0.649)}; \addlegendentry{DL 2020}
\addplot[mark=triangle*, thick, dotted] coordinates {(0,0.653) (4,0.708) (8,0.729) (12,0.747) (16,0.750)}; \addlegendentry{DL 2021}
\addplot[only marks, mark=o, mark size=2.2pt] coordinates {(3,0.612) (3,0.514) (3,0.663)}; \addlegendentry{BM25 top 3}
\addplot[only marks, mark=x, mark size=2.8pt, thick] coordinates {(16.9,0.573) (16.9,0.513) (16.9,0.687)}; \addlegendentry{Haiku, 16 labels}
\addplot[only marks, mark=star, mark size=2.8pt, thick] coordinates {(16.9,0.713)}; \addlegendentry{Sonnet list, 16}
\addplot[only marks, mark=+, mark size=2.8pt, thick] coordinates {(-0.4,0.581) (-0.4,0.524) (-0.4,0.617)}; \addlegendentry{Sonnet, no labels}
\addplot[mark=*, gray, thin, forget plot] coordinates {(4,0.582) (8,0.658) (12,0.685) (16,0.706)};
\addplot[mark=square*, gray, thin, forget plot] coordinates {(4,0.465) (8,0.491) (12,0.528) (16,0.543)};
\addplot[mark=triangle*, gray, thin, forget plot] coordinates {(4,0.513) (8,0.592) (12,0.640) (16,0.665)};
\addlegendimage{gray, thin}\addlegendentry{Topic classifier}
\end{axis}
\end{tikzpicture}
\caption{A few labels of the topic raise agreement. \methodL{} with 0--16 stratified labels and with the BM25 top 3 (open circles). Gray: the topic classifier.}
\label{fig:labels}
\end{figure}

\subsection{Topic labels}
\label{sec:labels}

A few human labels of a topic raise agreement substantially. Sixteen labels per topic raise the $\alpha$ of \methodL{} by 0.096 to 0.166 on the development and pilot collections. More than half of this gain comes from the first four labels, one per grade. The intervals of all these gains exclude 0. Figure~\ref{fig:labels} and Table~\ref{tab:pilot} in Appendix~\ref{app:more} show these results. The labels also remove most of the bias toward high grades. A likely reason is that the guidelines alone leave open where the assessors drew the lines between the grades. The labels supply this information.

The choice of labeled passages matters more than their number. The three BM25 top passages change $\alpha$ by at most 0.031. The intervals of these changes include 0. A likely reason is that the BM25 top passages do not cover all grades. Appendix~\ref{app:details} discusses which passages to label.

\methodL{} stays above the LLM judges when the judges receive the same labels. The labels raise the $\alpha$ of the judges as well. \methodL{} remains above the listwise Sonnet judge by 0.117 on DL 2020 and by 0.037 on DL 2021. The intervals of both differences exclude 0. Appendix~\ref{app:more} lists the differences from the other judges. With labels, the Sonnet judges still grade too high on DL 2021. A likely reason is that the writer turns the labels into explicit rules. The programs apply these rules in the same way to every document of the pool.

\subsection{Comparison with supervised per-topic methods}
\label{sec:cls}

\methodL{} agrees with NIST labels more than the topic classifier trained on the same labels. On the development and pilot collections, the gap in ordinal $\alpha$ is 0.050 to 0.106. The bootstrap intervals of all three gaps exclude 0. The paired t-test of Table~\ref{tab:main} is not significant on DL 2019. The gap grows with fewer labels. It also holds for binary $\alpha$. One-shot (16) is further below \methodL{}. Appendix~\ref{app:more} lists these values. The classifier grades too high. The analysis of the cut points below explains this bias.

The classifier needs many more labels to reach higher agreement than \methodL{}. With four times as many labels chosen by BM25, the classifier remains below \methodL{} with stratified labels. In the published setting of Gienapp et al., their classifiers use 64 to 256 labels per topic. These classifiers agree with NIST more than \methodL{} with 16 labels. Tables~\ref{tab:clsvar} and~\ref{tab:gienapp} in Appendix~\ref{app:more} list these results. The two methods therefore fit different label budgets. \methodL{} fits a few labels per topic. The classifier fits larger label sets.

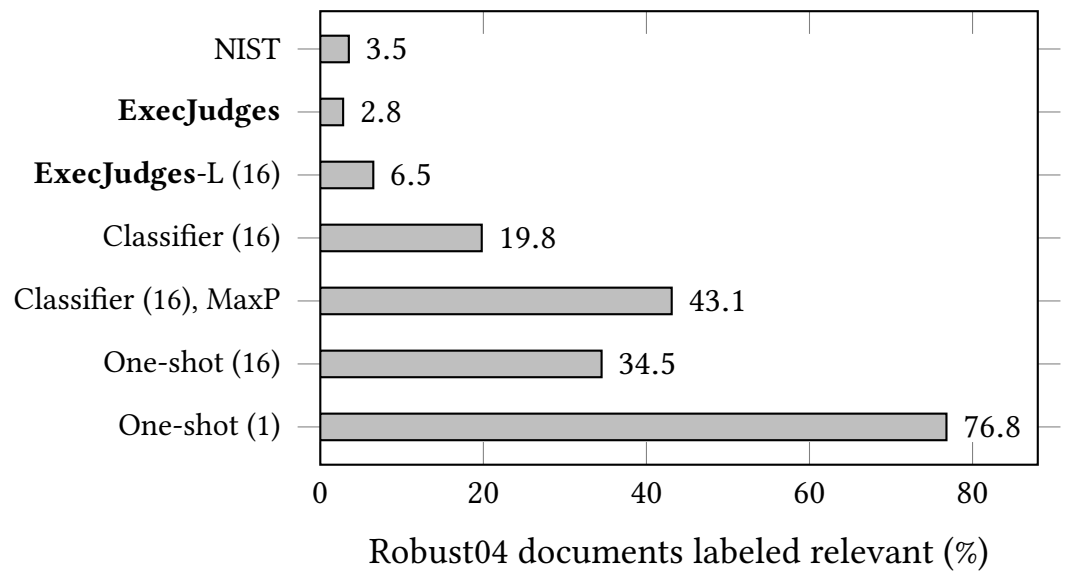
\begin{figure}[t]
\centering
\begin{tikzpicture}
\begin{axis}[xbar, scale only axis, width=0.56\columnwidth, height=3.0cm, bar width=5pt, xmin=0, xmax=88, xlabel={Robust04 documents labeled relevant (\%)}, ytick={0,1,2,3,4,5,6}, yticklabels={{NIST},{\method{}},{\methodL{} (16)},{Classifier (16)},{Classifier (16), MaxP},{One-shot (16)},{One-shot (1)}}, y dir=reverse, ymin=-0.6, ymax=6.6, nodes near coords, nodes near coords style={font=\scriptsize}, tick label style={font=\scriptsize}, label style={font=\footnotesize}]
\addplot[fill=gray!50, draw=black] coordinates {(3.5,0) (2.8,1) (6.5,2) (19.8,3) (43.1,4) (34.5,5) (76.8,6)};
\end{axis}
\end{tikzpicture}
\caption{Share of Robust04 evaluation documents labeled relevant. Methods fitted to the stratified labels over-predict relevance.}
\label{fig:robshare}
\end{figure}

\label{sec:heldout}\methodL{} also agrees with NIST labels more than the classifier and One-shot on the test collections. DL 2022, Robust04 and TREC-COVID were not used while designing the methods. \methodL{} is above the classifier on 69\% to 89\% of the topics of each collection. It is above One-shot (16) on a larger share of the topics. Figure~\ref{fig:pertopic} shows the per-topic results. The advantage of \methodL{} therefore does not depend on the collections that we used to design the method.

The gap is largest when relevant documents are rare, because methods fitted to the labels then label too many documents relevant. Only 3.5\% of the Robust04 evaluation documents are relevant. More than half of the stratified labels of a topic are relevant. Figure~\ref{fig:robshare} shows this effect. Most of the gap comes from the cut points that turn scores into grades. Oracle cut points give the documents of each topic exactly their NIST grade distribution. With oracle cut points, both supervised baselines improve and remain significantly below \methodL{} at $p<0.05$. Class weights, MaxP scoring, the mean of per-topic $\alpha$ and the removal of duplicate DL 2022 passages do not change this order. Table~\ref{tab:clsvar} in Appendix~\ref{app:more} and Appendix~\ref{app:details} report these checks.

\subsection{System rankings}
\label{sec:rank}

\begin{figure*}[t]
\centering
\begin{tikzpicture}
\begin{axis}[width=0.47\textwidth, height=5.2cm, title={\footnotesize DL 2021 (62 runs)}, xlabel={nDCG@10, NIST labels}, ylabel={nDCG@10, method labels}, xmin=0.2, xmax=0.95, ymin=0.2, ymax=0.95, legend style={font=\scriptsize, at={(0.02,0.98)}, anchor=north west, draw=none, fill=white, fill opacity=0.8, text opacity=1}, tick label style={font=\scriptsize}, label style={font=\footnotesize}, legend cell align=left]
\addplot[gray, dashed, forget plot] coordinates {(0.2,0.2) (0.95,0.95)};
\addplot[only marks, mark=*, mark size=1.3pt, black, mark options={fill opacity=0.6}] coordinates {(0.553,0.626) (0.513,0.591) (0.545,0.621) (0.672,0.716) (0.609,0.663) (0.736,0.776) (0.508,0.518) (0.566,0.561) (0.548,0.545) (0.455,0.496) (0.416,0.463) (0.642,0.700) (0.653,0.683) (0.522,0.602) (0.423,0.489) (0.603,0.636) (0.673,0.720) (0.583,0.621) (0.651,0.716) (0.626,0.679) (0.645,0.705) (0.642,0.702) (0.649,0.697) (0.636,0.693) (0.604,0.631) (0.678,0.729) (0.612,0.648) (0.448,0.529) (0.446,0.555) (0.660,0.692) (0.619,0.641) (0.637,0.651) (0.580,0.595) (0.594,0.615) (0.516,0.535) (0.465,0.463) (0.577,0.637) (0.752,0.807) (0.752,0.807) (0.752,0.807) (0.697,0.750) (0.709,0.768) (0.709,0.770) (0.645,0.676) (0.606,0.634) (0.539,0.627) (0.406,0.424) (0.396,0.413) (0.524,0.551) (0.465,0.566) (0.476,0.575) (0.463,0.489) (0.649,0.705) (0.374,0.434) (0.452,0.522) (0.495,0.575) (0.479,0.554) (0.522,0.608) (0.549,0.622) (0.697,0.731) (0.573,0.627) (0.693,0.728)};
\addlegendentry{\method{}, $\tau$ 0.869}
\addplot[only marks, mark=triangle*, mark size=1.3pt, blue!70!black, mark options={fill opacity=0.6}] coordinates {(0.553,0.550) (0.513,0.513) (0.545,0.539) (0.672,0.656) (0.609,0.621) (0.736,0.723) (0.508,0.489) (0.566,0.537) (0.548,0.510) (0.455,0.454) (0.416,0.424) (0.642,0.643) (0.653,0.642) (0.522,0.531) (0.423,0.437) (0.603,0.605) (0.673,0.661) (0.583,0.561) (0.651,0.657) (0.626,0.623) (0.645,0.629) (0.642,0.634) (0.649,0.637) (0.636,0.628) (0.604,0.593) (0.678,0.673) (0.612,0.595) (0.448,0.455) (0.446,0.472) (0.660,0.650) (0.619,0.595) (0.637,0.616) (0.580,0.574) (0.594,0.593) (0.516,0.511) (0.465,0.443) (0.577,0.582) (0.752,0.744) (0.752,0.744) (0.752,0.744) (0.697,0.686) (0.709,0.700) (0.709,0.701) (0.645,0.628) (0.606,0.597) (0.539,0.543) (0.406,0.397) (0.396,0.382) (0.524,0.510) (0.465,0.487) (0.476,0.502) (0.463,0.479) (0.649,0.645) (0.374,0.390) (0.452,0.467) (0.495,0.530) (0.479,0.503) (0.522,0.544) (0.549,0.558) (0.697,0.674) (0.573,0.562) (0.693,0.679)};
\addlegendentry{\methodL{} (16), $\tau$ 0.926}
\addplot[only marks, mark=x, mark size=1.3pt, red!70!black, mark options={fill opacity=0.6}] coordinates {(0.553,0.656) (0.513,0.609) (0.545,0.649) (0.672,0.826) (0.609,0.758) (0.736,0.875) (0.508,0.665) (0.566,0.734) (0.548,0.720) (0.455,0.534) (0.416,0.501) (0.642,0.775) (0.653,0.785) (0.522,0.650) (0.423,0.525) (0.603,0.772) (0.673,0.799) (0.583,0.682) (0.651,0.775) (0.626,0.733) (0.645,0.768) (0.642,0.762) (0.649,0.767) (0.636,0.751) (0.604,0.738) (0.678,0.797) (0.612,0.746) (0.448,0.538) (0.446,0.548) (0.660,0.793) (0.619,0.748) (0.637,0.761) (0.580,0.712) (0.594,0.728) (0.516,0.668) (0.465,0.626) (0.577,0.718) (0.752,0.890) (0.752,0.890) (0.752,0.890) (0.697,0.811) (0.709,0.825) (0.709,0.825) (0.645,0.767) (0.606,0.719) (0.539,0.640) (0.406,0.520) (0.396,0.506) (0.524,0.625) (0.465,0.560) (0.476,0.576) (0.463,0.570) (0.649,0.777) (0.374,0.528) (0.452,0.587) (0.495,0.606) (0.479,0.589) (0.522,0.624) (0.549,0.640) (0.697,0.825) (0.573,0.714) (0.693,0.821)};
\addlegendentry{Sonnet, pointwise, $\tau$ 0.896}
\end{axis}
\end{tikzpicture}\hfill
\begin{tikzpicture}
\begin{axis}[width=0.47\textwidth, height=5.2cm, title={\footnotesize DL 2022 (100 runs)}, xlabel={nDCG@10, NIST labels}, ylabel={nDCG@10, method labels}, xmin=0.2, xmax=0.95, ymin=0.2, ymax=0.95, legend style={font=\scriptsize, at={(0.02,0.98)}, anchor=north west, draw=none, fill=white, fill opacity=0.8, text opacity=1}, tick label style={font=\scriptsize}, label style={font=\footnotesize}, legend cell align=left]
\addplot[gray, dashed, forget plot] coordinates {(0.2,0.2) (0.95,0.95)};
\addplot[only marks, mark=*, mark size=1.3pt, black, mark options={fill opacity=0.6}] coordinates {(0.599,0.605) (0.574,0.606) (0.575,0.608) (0.414,0.456) (0.362,0.399) (0.439,0.498) (0.463,0.506) (0.482,0.515) (0.481,0.504) (0.528,0.553) (0.526,0.558) (0.710,0.689) (0.710,0.691) (0.704,0.692) (0.514,0.546) (0.553,0.565) (0.532,0.550) (0.511,0.522) (0.550,0.558) (0.568,0.584) (0.590,0.596) (0.568,0.578) (0.585,0.595) (0.562,0.577) (0.581,0.586) (0.567,0.588) (0.505,0.549) (0.521,0.488) (0.486,0.532) (0.591,0.594) (0.481,0.525) (0.591,0.598) (0.433,0.466) (0.495,0.528) (0.464,0.497) (0.707,0.693) (0.507,0.527) (0.507,0.531) (0.548,0.580) (0.570,0.606) (0.497,0.519) (0.531,0.547) (0.264,0.355) (0.260,0.351) (0.261,0.350) (0.357,0.433) (0.360,0.444) (0.362,0.439) (0.702,0.705) (0.628,0.631) (0.550,0.561) (0.536,0.549) (0.457,0.516) (0.479,0.537) (0.397,0.446) (0.407,0.457) (0.699,0.675) (0.707,0.680) (0.712,0.682) (0.277,0.312) (0.257,0.286) (0.259,0.297) (0.365,0.409) (0.367,0.403) (0.374,0.406) (0.576,0.601) (0.597,0.613) (0.288,0.385) (0.319,0.406) (0.310,0.400) (0.311,0.403) (0.660,0.679) (0.645,0.667) (0.647,0.670) (0.567,0.602) (0.633,0.656) (0.657,0.670) (0.542,0.448) (0.527,0.459) (0.587,0.600) (0.621,0.632) (0.641,0.631) (0.545,0.563) (0.536,0.560) (0.284,0.372) (0.293,0.383) (0.621,0.623) (0.650,0.633) (0.561,0.576) (0.567,0.608) (0.651,0.634) (0.650,0.634) (0.650,0.634) (0.513,0.523) (0.497,0.520) (0.495,0.522) (0.490,0.518) (0.530,0.542) (0.605,0.609) (0.505,0.515)};
\addlegendentry{\method{}, $\tau$ 0.882}
\addplot[only marks, mark=triangle*, mark size=1.3pt, blue!70!black, mark options={fill opacity=0.6}] coordinates {(0.599,0.601) (0.574,0.585) (0.575,0.589) (0.414,0.431) (0.362,0.398) (0.439,0.461) (0.463,0.482) (0.482,0.492) (0.481,0.482) (0.528,0.524) (0.526,0.527) (0.710,0.678) (0.710,0.678) (0.704,0.676) (0.514,0.520) (0.553,0.550) (0.532,0.524) (0.511,0.510) (0.550,0.545) (0.568,0.564) (0.590,0.583) (0.568,0.559) (0.585,0.580) (0.562,0.569) (0.581,0.582) (0.567,0.572) (0.505,0.508) (0.521,0.496) (0.486,0.493) (0.591,0.581) (0.481,0.488) (0.591,0.586) (0.433,0.440) (0.495,0.500) (0.464,0.483) (0.707,0.677) (0.507,0.500) (0.507,0.504) (0.548,0.555) (0.570,0.570) (0.497,0.494) (0.531,0.523) (0.264,0.321) (0.260,0.321) (0.261,0.322) (0.357,0.420) (0.360,0.428) (0.362,0.435) (0.702,0.681) (0.628,0.606) (0.550,0.541) (0.536,0.543) (0.457,0.472) (0.479,0.492) (0.397,0.401) (0.407,0.414) (0.699,0.665) (0.707,0.676) (0.712,0.679) (0.277,0.284) (0.257,0.270) (0.259,0.278) (0.365,0.403) (0.367,0.409) (0.374,0.416) (0.576,0.574) (0.597,0.591) (0.288,0.340) (0.319,0.371) (0.310,0.366) (0.311,0.367) (0.660,0.650) (0.645,0.637) (0.647,0.641) (0.567,0.572) (0.633,0.627) (0.657,0.639) (0.542,0.486) (0.527,0.488) (0.587,0.585) (0.621,0.617) (0.641,0.628) (0.545,0.545) (0.536,0.537) (0.284,0.327) (0.293,0.335) (0.621,0.618) (0.650,0.628) (0.561,0.567) (0.567,0.600) (0.651,0.630) (0.650,0.629) (0.650,0.629) (0.513,0.526) (0.497,0.494) (0.495,0.502) (0.490,0.494) (0.530,0.523) (0.605,0.595) (0.505,0.492)};
\addlegendentry{\methodL{} (16), $\tau$ 0.925}
\addplot[only marks, mark=x, mark size=1.3pt, red!70!black, mark options={fill opacity=0.6}] coordinates {(0.599,0.884) (0.574,0.853) (0.575,0.858) (0.414,0.639) (0.362,0.525) (0.439,0.656) (0.463,0.713) (0.482,0.688) (0.481,0.695) (0.528,0.741) (0.526,0.741) (0.710,0.900) (0.710,0.901) (0.704,0.897) (0.514,0.697) (0.553,0.805) (0.532,0.758) (0.511,0.726) (0.550,0.771) (0.568,0.808) (0.590,0.809) (0.568,0.800) (0.585,0.797) (0.562,0.807) (0.581,0.802) (0.567,0.865) (0.505,0.713) (0.521,0.693) (0.486,0.711) (0.591,0.840) (0.481,0.676) (0.591,0.841) (0.433,0.615) (0.495,0.708) (0.464,0.669) (0.707,0.898) (0.507,0.738) (0.507,0.737) (0.548,0.782) (0.570,0.817) (0.497,0.721) (0.531,0.778) (0.264,0.398) (0.260,0.363) (0.261,0.379) (0.357,0.582) (0.360,0.565) (0.362,0.576) (0.702,0.900) (0.628,0.859) (0.550,0.802) (0.536,0.772) (0.457,0.726) (0.479,0.704) (0.397,0.613) (0.407,0.603) (0.699,0.889) (0.707,0.897) (0.712,0.894) (0.277,0.338) (0.257,0.304) (0.259,0.311) (0.365,0.546) (0.367,0.512) (0.374,0.519) (0.576,0.807) (0.597,0.847) (0.288,0.413) (0.319,0.453) (0.310,0.444) (0.311,0.445) (0.660,0.844) (0.645,0.827) (0.647,0.833) (0.567,0.783) (0.633,0.842) (0.657,0.878) (0.542,0.704) (0.527,0.693) (0.587,0.848) (0.621,0.840) (0.641,0.847) (0.545,0.774) (0.536,0.717) (0.284,0.404) (0.293,0.407) (0.621,0.881) (0.650,0.867) (0.561,0.806) (0.567,0.792) (0.651,0.867) (0.650,0.867) (0.650,0.867) (0.513,0.709) (0.497,0.682) (0.495,0.691) (0.490,0.680) (0.530,0.715) (0.605,0.838) (0.505,0.709)};
\addlegendentry{Classifier (16), $\tau$ 0.866}
\end{axis}
\end{tikzpicture}
\caption{System rankings in TREC style. Each point is one official run: its nDCG@10 under NIST labels and under a method's labels. Points on the diagonal have the same score under both. Labeled passages keep their NIST grades. DL 2022: 65 topics.}
\label{fig:runs}
\end{figure*}
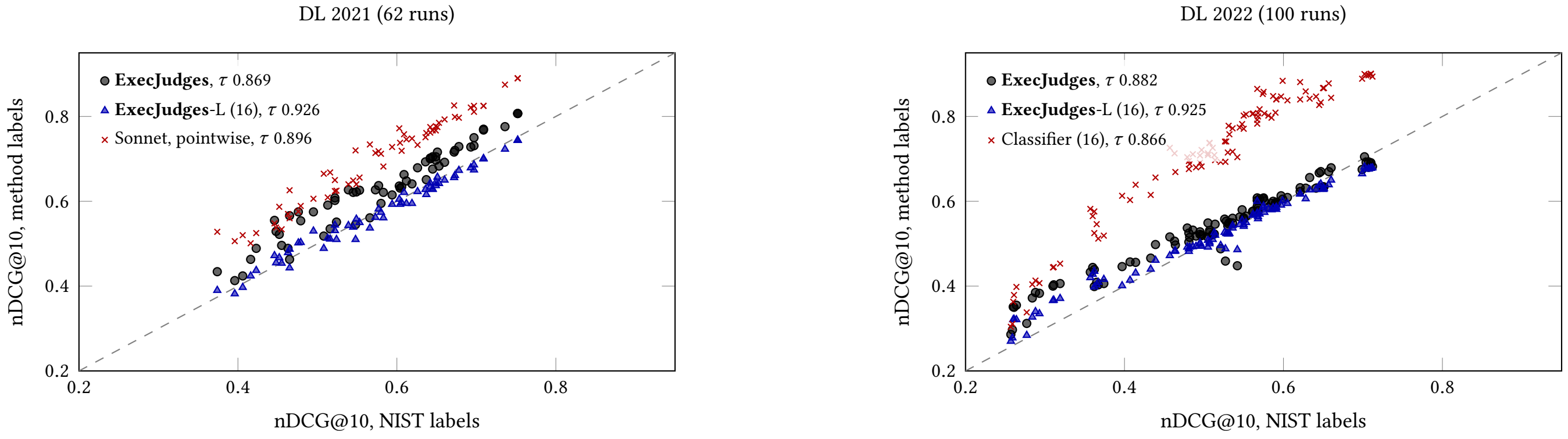

\method{} and \methodL{} rank the official runs close to the ranking under NIST labels. We rank the official passage runs of DL 2019--2022 by nDCG@10 under each label set. Labeled passages keep their NIST grades. Unjudged passages have gain 0. Over all runs, Kendall's $\tau$ is 0.87 to 0.91 for \method{} and 0.93 to 0.94 for \methodL{} with 16 labels. Figure~\ref{fig:runs} shows DL 2021 and 2022. Table~\ref{tab:tau-dl} in Appendix~\ref{app:more} lists all label sets. Ranking agreement at this level is usually considered acceptable for comparing systems.

The topic classifier ranks the runs less accurately and inflates their scores. With the same labels, its $\tau$ is 0.059 to 0.085 below \methodL{} on DL 2020--2022. The intervals of these differences exclude 0. The classifier labels also raise the nDCG@10 of every run. The \methodL{} labels change nDCG@10 by less than 0.01 on average. This inflation follows from the over-grading of the classifier reported in Section~\ref{sec:cls}.

Label agreement and ranking agreement can differ. The embedding cutoffs have low $\alpha$ on DL 2019 and 2020. Their $\tau$ is still above 0.85. The BM25 cutoffs give negative $\tau$. These cutoffs give the highest gains to the passages that BM25-like runs retrieve. A label source therefore has to be evaluated with both measures.

\methodL{} labels also make the pools the most reusable among the label sources we tested. We leave one run group at a time out of the DL 2019--2021 pools. The judged passages that only this group retrieved are then counted as non-relevant or labeled by a method. \methodL{} labels change the nDCG@10 of the left-out runs by less than 0.005 on average. This change is significantly smaller than for every other label source. Table~\ref{tab:holes} in Appendix~\ref{app:more} lists all label sources. Program labels can therefore fill the unjudged passages of a pool when new systems are evaluated.

\subsection{Cost}
\label{sec:cost}

\begin{figure}[t]
\centering
\begin{tikzpicture}
\begin{axis}[width=\columnwidth, height=3.8cm, xlabel={Documents judged per topic}, ylabel={API cost per topic (USD)}, xmin=0, xmax=1000, ymin=0, ymax=1.2, legend style={font=\scriptsize, at={(0.02,0.98)}, anchor=north west, draw=none, fill=white, fill opacity=0.8, text opacity=1}, tick label style={font=\scriptsize}, label style={font=\footnotesize}, legend cell align=left]
\addplot[thick, blue!70!black] coordinates {(0,0.21) (1000,0.21)}; \addlegendentry{\methodL{}, DL}
\addplot[thick, blue!70!black, dashed] coordinates {(0,0.40) (1000,0.40)}; \addlegendentry{\methodL{}, Robust04 / TREC-COVID}
\addplot[thick, red!70!black] coordinates {(0,0) (1000,0.60)}; \addlegendentry{Sonnet, listwise, DL}
\addplot[thick, red!70!black, dashed] coordinates {(0,0) (600,1.2)}; \addlegendentry{Sonnet, listwise, Robust04 / TREC-COVID}
\addplot[only marks, mark=*, mark size=1.5pt, black, forget plot] coordinates {(350,0.21) (200,0.40)};
\end{axis}
\end{tikzpicture}
\caption{API cost per topic with 16 labels. Writing the programs is a fixed cost. The programs are cheaper than the listwise judge beyond about 200 to 350 documents per topic (dots).}
\label{fig:cost}
\end{figure}
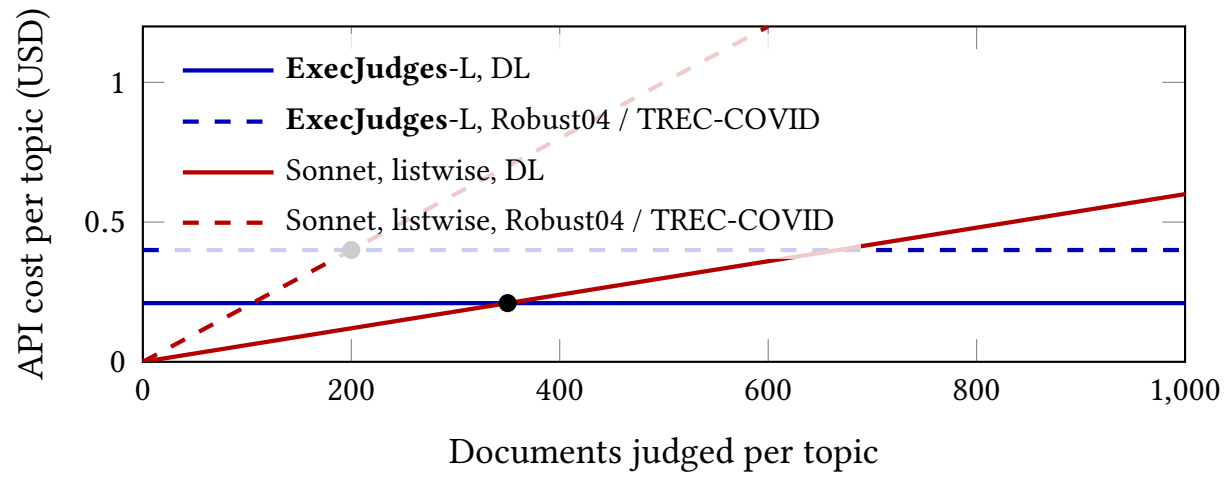

The programs are cheaper than the listwise Sonnet judge once a topic has more than about 200 to 350 documents. Figure~\ref{fig:cost} shows this break-even point. Writing the programs costs \$0.17 to \$0.49 per topic. The programs then grade any number of documents without API calls. Grading takes less than one second per topic. The cost of the programs is paid once per topic and does not depend on the number of documents. The cost of an LLM judge grows with every document. The saving therefore grows with the depth of the pool.

The supervised baselines trade API cost for GPU time. The topic classifier has no API cost. Its training takes 13 seconds per topic on a GPU. The programs need no GPU, because they use only deterministic, lightweight operations. Table~\ref{tab:cost} in Appendix~\ref{app:more} lists the cost and time per collection.

\section{Analysis}
\label{sec:analysis}

We now analyze what the programs do, which design choices matter, and where the programs fail.

\subsection{The generated programs}

The programs are short and mostly lexical. The median program has about 70 lines and two functions. Topic-specific regular expressions and term lists make most decisions. On TREC DL, nearly all programs call \texttt{jk.unrelated\_share} to decide whether a passage is dedicated to the query. On Robust04 and TREC-COVID, the programs encode the conditions of the narrative. For example, a whale-watching article counts only if it names Californian places at least half as often as other places. A hydroxychloroquine article gets grade 2 only if it describes a study. Figure~\ref{fig:programs} shows parts of these two programs. Table~\ref{tab:progstats} in Appendix~\ref{app:more} gives the statistics per collection. The topic-specific criteria are therefore explicit in the code.

\begin{figure*}[t]
\centering
\begin{minipage}[t]{0.48\textwidth}
\textbf{Robust04 660}: \emph{whale watching California}; narrative: \emph{mention of one or more Californian site or guide services for watching whales}
\scriptsize
\begin{lstlisting}[frame=single, breaklines=false, xleftmargin=3pt, xrightmargin=3pt]
WW_RE = re.compile(r"whale[\s-]*watch\w*|
  whale\s+sightings?|whale[\s-]*(tours?|
  trips?|cruises?)|...")
CA_PLACES = ["california", "monterey",
  "point reyes", "dana point", ...]  # 60
FOREIGN_RE = re.compile(r"\b(baja|mexico|
  hawaii|alaska|cape cod|iceland|...)\b")
SERVICE_RE = re.compile(r"\b(boats?|
  charters?|naturalists?|tickets?|...)\b")
def assess(query, results):
  for r in results:
    ca_ok = ca > 0 and ca >= foreign * 0.5
    relevant = (ww >= 1 and ca_ok) or
      (migr >= 1 and whales >= 2 and ca_ok
       and service >= 2)
    det = jk.detail(DETAIL_TERMS, text,
                    min_terms=2)
    key = relevant and ww >= 2 and det >= 4
          and service >= 4 and ca >= 2
\end{lstlisting}
\end{minipage}\hfill
\begin{minipage}[t]{0.48\textwidth}
\textbf{TREC-COVID 28}: \emph{coronavirus hydroxychloroquine}; narrative: \emph{basic science or clinical studies assessing the benefit and harms of treating Covid-19 with hydroxychloroquine}
\scriptsize
\begin{lstlisting}[frame=single, breaklines=false, xleftmargin=3pt, xrightmargin=3pt]
HCQ_PAT = re.compile(r"hydroxy\s*-?\s*
  chloroquine|\bhcq\b|plaquenil|...")
EVIDENCE_PAT = re.compile(r"randomi[sz]ed|
  clinical trial|cohort|in vitro|ec50|
  mortality|viral load|qt prolong|...")
STUDY_PAT = re.compile(r"randomi[sz]ed|
  cohort|meta-analysis|in vitro|\d+ patients|
  confidence interval|95%|...")
def assess(query, results):
  for r in results:
    hcq_ev_sents = sum(1 for s in
      jk.sentences(full) if HCQ_PAT.search(s)
      and EVIDENCE_PAT.search(s))
    relevant = covid and (title_hcq or
      (n_hcq >= 2 and (hcq_ev_sents >= 1
                       or det >= 2)) or ...)
    key = relevant and title_hcq and
          study >= 1 and hcq_ev_sents >= 1
\end{lstlisting}
\end{minipage}
\caption{Parts of two \method{} programs, shortened (\texttt{...}). The two answers encode the narrative of the topic. Figure~\ref{fig:example} shows a TREC DL program.}
\label{fig:programs}
\end{figure*}

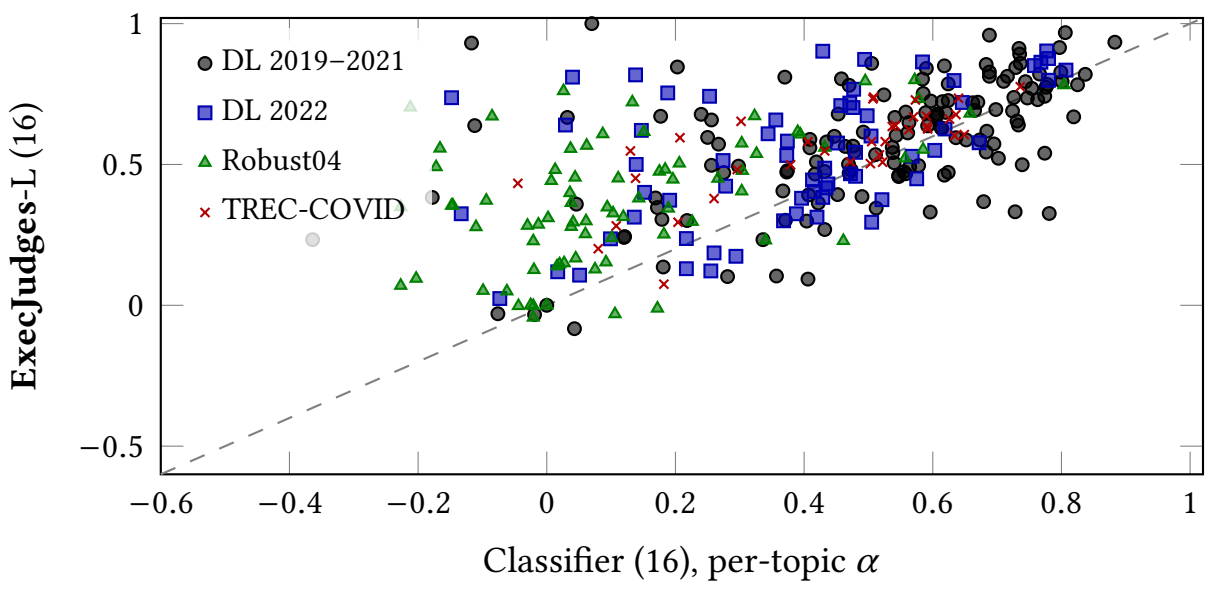
\begin{figure}[t]
\centering
\begin{tikzpicture}
\begin{axis}[width=\columnwidth, height=4.6cm, xlabel={Classifier (16), per-topic $\alpha$}, ylabel={\methodL{} (16)}, xmin=-0.6, xmax=1.02, ymin=-0.6, ymax=1.02, legend style={font=\scriptsize, at={(0.02,0.98)}, anchor=north west, draw=none, fill=white, fill opacity=0.8, text opacity=1}, tick label style={font=\scriptsize}, label style={font=\footnotesize}, legend cell align=left]
\addplot[gray, dashed, forget plot] coordinates {(-0.6,-0.6) (1.02,1.02)};
\addplot[only marks, mark=*, mark size=1.2pt, black, mark options={fill opacity=0.6}] coordinates {(0.070,1.000) (0.763,0.730) (0.409,0.589) (0.298,0.494) (0.555,0.474) (0.374,0.478) (0.683,0.544) (0.778,0.784) (0.181,0.136) (0.683,0.855) (0.806,0.968) (0.744,0.794) (0.883,0.934) (0.774,0.540) (0.797,0.915) (0.469,0.501) (0.537,0.561) (0.564,0.492) (0.404,0.299) (0.825,0.783) (0.662,0.719) (0.605,0.668) (0.120,0.241) (0.505,0.859) (0.774,0.742) (0.447,0.600) (0.547,0.457) (0.203,0.846) (0.596,0.331) (0.367,0.406) (0.511,0.535) (0.540,0.506) (0.490,0.386) (0.624,0.472) (0.837,0.820) (0.747,0.736) (0.547,0.461) (0.688,0.813) (0.766,0.821) (0.000,0.000) (0.710,0.795) (0.470,0.781) (-0.117,0.931) (0.240,0.678) (0.665,0.692) (0.538,0.542) (0.618,0.462) (0.620,0.685) (0.373,0.473) (0.275,0.471) (0.357,0.104) (-0.178,0.383) (0.179,0.305) (0.624,0.786) (0.492,0.615) (0.728,0.332) (-0.019,-0.033) (0.754,0.770) (0.679,0.368) (0.406,0.093) (0.043,-0.083) (0.589,0.684) (0.473,0.481) (0.464,0.564) (0.478,0.545) (0.512,0.345) (0.336,0.233) (0.172,0.347) (0.776,0.774) (0.615,0.723) (0.177,0.671) (0.781,0.326) (0.624,0.727) (0.432,0.269) (-0.076,-0.030) (0.046,0.359) (0.121,0.244) (0.561,0.612) (0.558,0.467) (0.726,0.738) (0.218,0.301) (0.256,0.658) (0.267,0.572) (0.590,0.841) (0.371,0.302) (-0.112,0.638) (0.734,0.912) (0.607,0.679) (0.724,0.690) (0.584,0.803) (0.541,0.667) (0.735,0.892) (0.613,0.630) (0.652,0.587) (0.801,0.794) (0.281,0.102) (0.688,0.828) (0.558,0.686) (0.250,0.596) (0.731,0.654) (0.543,0.603) (0.670,0.721) (0.419,0.509) (0.578,0.496) (0.702,0.522) (0.476,0.565) (0.032,0.667) (0.673,0.587) (0.256,0.497) (0.422,0.363) (0.370,0.810) (0.452,0.393) (0.453,0.679) (0.698,0.695) (0.417,0.466) (-0.364,0.233) (0.733,0.640) (0.169,0.381) (0.614,0.627) (0.819,0.669) (0.592,0.637) (0.695,0.573) (0.684,0.618) (0.799,0.828) (0.585,0.752) (0.432,0.475) (0.736,0.859) (0.716,0.813) (0.409,0.559) (0.563,0.649) (0.688,0.959) (0.605,0.673) (0.524,0.747) (0.636,0.595) (0.597,0.725) (0.729,0.844) (0.435,0.580) (0.470,0.470) (0.618,0.850) (0.739,0.499) (0.407,0.392) (0.458,0.804)};
\addlegendentry{DL 2019--2021}
\addplot[only marks, mark=square*, mark size=1.2pt, blue!70!black, mark options={fill opacity=0.6}] coordinates {(0.521,0.375) (0.505,0.295) (0.255,0.122) (0.575,0.449) (0.260,0.186) (0.452,0.576) (0.017,0.119) (0.429,0.383) (0.099,0.236) (0.480,0.458) (0.434,0.417) (0.646,0.720) (0.429,0.902) (0.477,0.767) (0.457,0.709) (0.494,0.873) (0.437,0.431) (0.188,0.754) (0.274,0.514) (-0.073,0.024) (0.672,0.577) (0.472,0.467) (0.368,0.300) (0.147,0.621) (0.432,0.487) (0.138,0.818) (0.619,0.626) (0.603,0.550) (0.344,0.609) (0.780,0.799) (0.139,0.500) (0.278,0.423) (0.584,0.864) (0.217,0.130) (-0.148,0.737) (0.758,0.851) (0.029,0.640) (0.420,0.313) (0.253,0.742) (-0.133,0.325) (0.294,0.174) (0.471,0.718) (0.768,0.863) (0.416,0.404) (0.476,0.702) (0.498,0.673) (0.051,0.107) (0.780,0.876) (0.777,0.903) (0.480,0.543) (0.136,0.313) (0.807,0.835) (0.152,0.401) (0.413,0.445) (0.633,0.798) (0.040,0.810) (0.356,0.658) (0.191,0.373) (0.373,0.532) (0.568,0.529) (0.217,0.237) (0.396,0.380) (0.504,0.601) (0.374,0.583) (0.388,0.325)};
\addlegendentry{DL 2022}
\addplot[only marks, mark=triangle*, mark size=1.2pt, green!50!black, mark options={fill opacity=0.6}] coordinates {(-0.021,-0.002) (0.461,0.229) (-0.021,0.228) (-0.021,-0.043) (0.060,0.253) (0.037,0.556) (0.151,0.616) (0.206,0.504) (-0.166,0.557) (0.394,0.607) (0.007,0.442) (0.133,0.721) (-0.172,0.491) (0.189,0.344) (0.326,0.538) (0.495,0.796) (0.196,0.447) (-0.025,0.001) (0.027,0.150) (0.026,0.761) (0.036,0.399) (-0.099,0.052) (0.119,0.314) (-0.000,0.000) (0.303,0.405) (0.087,0.608) (0.572,0.800) (0.016,0.139) (-0.212,0.703) (0.092,0.153) (0.002,0.310) (0.172,-0.011) (-0.094,0.372) (0.062,0.567) (0.044,0.454) (0.304,0.478) (-0.030,0.283) (0.013,0.481) (-0.085,0.671) (-0.062,0.050) (0.038,0.364) (0.578,0.733) (0.585,0.553) (-0.044,-0.002) (0.804,0.782) (-0.020,0.126) (-0.148,0.354) (0.038,0.279) (0.182,0.252) (0.175,0.475) (0.111,0.451) (0.266,0.450) (0.101,0.240) (-0.227,0.349) (0.020,0.146) (0.340,0.231) (0.658,0.681) (0.106,-0.031) (0.094,0.348) (0.041,0.295) (0.184,0.482) (0.323,0.672) (0.045,0.167) (0.303,0.575) (-0.146,0.355) (0.075,0.128) (-0.110,0.278) (0.557,0.522) (0.103,0.327) (-0.203,0.095) (-0.013,0.287) (0.061,0.299) (0.226,0.297) (0.432,0.560) (0.143,0.380) (-0.227,0.070) (0.390,0.616)};
\addlegendentry{Robust04}
\addplot[only marks, mark=x, mark size=1.2pt, red!70!black, mark options={fill opacity=0.6}] coordinates {(0.207,0.595) (0.182,0.075) (0.517,0.534) (-0.045,0.433) (0.592,0.626) (0.503,0.503) (0.204,0.295) (0.573,0.729) (0.508,0.739) (0.736,0.776) (0.406,0.581) (0.570,0.669) (0.507,0.733) (0.505,0.581) (0.591,0.636) (0.640,0.735) (0.634,0.602) (0.260,0.379) (0.130,0.548) (0.108,0.282) (0.138,0.451) (0.522,0.508) (0.472,0.509) (0.526,0.581) (0.563,0.624) (0.626,0.661) (0.542,0.638) (0.537,0.635) (0.302,0.653) (0.590,0.672) (0.432,0.548) (0.080,0.201) (0.649,0.606) (0.636,0.677) (0.379,0.499) (0.296,0.482)};
\addlegendentry{TREC-COVID}
\end{axis}
\end{tikzpicture}
\caption{\methodL{} is above the topic classifier on most topics. Per-topic ordinal $\alpha$ with the same 16 labels. Points above the diagonal favor \methodL{}.}
\label{fig:pertopic}
\end{figure}

\methodL{} agrees with NIST more than the classifier on most topics. With the same labels, \methodL{} is ahead on 84 of the 142 development and pilot topics. A sign test gives $p<0.001$. Figure~\ref{fig:pertopic} shows the per-topic values. The largest gains are on topics about a named entity, such as \emph{who sings monk theme song}, \emph{who is robert gray} and \emph{what is mamey}. On these topics, a relevant passage has to name the entity. The classifier is ahead on topics whose answer is a number or a comparison, such as \emph{reba mcentire's net worth} and \emph{difference between a hotel and motel}. The rules of the programs do not capture these answers. The difference between the two methods is not related to pool size, the share of relevant passages or query length.

\subsection{Design choices}
\label{sec:devpilot}

The programs have to be written for each topic to reach the agreement of the LLM judges. The lexical classifier learns from Sonnet labels with the same kind of features as the programs. It is 0.20 to 0.47 below \method{} on the development and pilot collections. \method{}-Global and \method{}-GR share one program or one rubric across all topics. Both variants are below the lexical classifier on every year. Table~\ref{tab:pilot} in Appendix~\ref{app:more} lists these results. The topic-specific rules and term lists therefore account for most of the agreement.

Averaging more programs improves agreement up to three programs. We wrote two more programs per topic from the same prompt on the development and pilot collections. We then averaged every subset of $k$ programs. From $k=1$ to $k=3$, $\alpha$ rises by 0.016 to 0.031. Most of the gain comes from the second program. Beyond three programs, the gain is smaller than the difference between two rewrites of the same program. Each program adds one writer call per topic. We therefore fixed $k=3$.

\method{} depends on a strong program writer. We replaced the program writer with an open-weight model and kept the same prompts, tests and repair step. Our implementation uses GPT-OSS-120B~\cite{openai2025gptoss} as the open-weight writer. With this writer, the $\alpha$ of \method{} is at most 0.486 on the development and pilot collections. All values of \method{} and \methodL{} are significantly below the values with the main writer at $p<0.01$. Table~\ref{tab:pilot} lists these results. The programs of the open-weight writer grade too low. Writing rules that match the criteria of the assessors therefore requires a capable writer model at present.

Corrections based on labeled errors did not raise agreement on new passages. We tried four ways to correct program errors after we fixed the main method. None of them raised agreement on passages outside those used for the correction. Appendix~\ref{app:more} describes these attempts. A likely reason is that corrections fitted to a few labeled errors do not generalize to other passages.

\subsection{Errors}

The largest error of \method{} is a missed top grade. On DL 2019 and 2020, \method{} gives grade 0 to more than a quarter of the passages that NIST graded 3. A few topics hold most of these errors. The labels remove most of this error. With labels, the share falls below 9\%. \methodL{} also over-grades less than the classifier with the same labels. Table~\ref{tab:errors} in Appendix~\ref{app:more} lists all error types. A likely reason is that a labeled passage of the top grade shows the writer which evidence the assessors accepted for that grade.

\begin{table}[t]
\caption{DL 2022 passages with large errors. N: NIST; E: \method{}; L: \methodL{} (16); S: Sonnet, pointwise; C: classifier (16).}
\label{tab:errex}
\footnotesize
\setlength{\tabcolsep}{2.2pt}
\begin{adjustbox}{max width=\columnwidth}
\begin{tabular}{@{}p{0.78\columnwidth}ccccc@{}}
\toprule
Query: passage & N & E & L & S & C \\
\midrule
\emph{what is cold bridging}: ``Think of anything in your wall that's not insulation as a bridge. Heat and cold will travel across that bridge\ldots'' & 2 & 0 & 0 & 2 & 1 \\
\multicolumn{6}{@{}p{0.98\columnwidth}@{}}{\scriptsize Every program looks for the phrase \emph{cold}, \emph{heat} or \emph{thermal bridge}. The passage contains none of them.} \\[2pt]
\emph{example of what a family advocate does}: ``Some of the duties of a Child advocate might include: Assessment of child and family\ldots{} Practical assistance to family.'' & 2 & 0 & 2 & 2 & 2 \\
\multicolumn{6}{@{}p{0.98\columnwidth}@{}}{\scriptsize The programs without labels give grade 2 mainly to passages that name family or parent advocates. The labels include victim and nurse advocates graded 2, and the programs written with them accept other advocates who work with families or children.} \\[2pt]
\emph{what does london breed stand for}: ``London is a Certified Applied Animal Behaviorist and Certified Pet Dog Trainer\ldots'' & 0 & 0 & 0 & 0 & 2 \\
\multicolumn{6}{@{}p{0.98\columnwidth}@{}}{\scriptsize The query names the politician London Breed. The passage is about a dog trainer named London.} \\[2pt]
\emph{what is the name of a baby nurse}: ``\ldots{}the meaning of Fatima is `baby's nurse'.'' & 0 & 0 & 0 & 2 & 0 \\
\multicolumn{6}{@{}p{0.98\columnwidth}@{}}{\scriptsize The query words appear as the meaning of a first name.} \\
\bottomrule
\end{tabular}
\end{adjustbox}
\end{table}

The four methods make different kinds of errors. We compare them on the DL 2022 passages that the pointwise Sonnet judge also labeled. The classifier with 16 labels is two or more grades off the NIST grade on 17.0\% of these passages. \method{}, \methodL{} and the Sonnet judge are two or more grades off on at most 7.7\%. Most errors of the classifier are grades of 2 or 3 for passages that NIST graded 0. \method{} makes the opposite error. It gives grade 0 to passages that NIST graded 2 or 3. Table~\ref{tab:errex} shows one example of each kind of error. Appendix~\ref{app:errors} gives the error counts of each method. These patterns match the biases reported above. The classifier grades too high. The programs without labels miss relevant passages.

\subsection{Implications and limitations}
\label{sec:limits}

The results indicate where each variant fits. \method{} without labels fits passage collections with short queries. On the development and pilot collections, it agreed with NIST at the level of the Sonnet judges. \methodL{} fits when an assessor can label a few passages per topic. With 16 labels, it agreed with NIST more than both supervised baselines on every collection. Two settings call for caution: document collections with narrative topics, and the ordering of the best runs. The cost of both variants is fixed per topic. The saving over an LLM judge therefore grows with the depth of the pool.

Our study has four limitations. They concern the human labels, the ranking of the best systems, evaluation gaming, and the models that we used. For each limitation, we say what it means for the results.

\paragraph{Dependence on human labels.} The better-performing variant, \methodL{}, requires human labels of each topic. In our experiments, the labels are stratified by grade. This selection requires the assessor grades before labeling. A practical procedure has to choose the passages without knowing their grades. The gains that we report for \methodL{} therefore show how much a given number of labels can help. A possible future direction is active learning, which requests labels from a human assessor or from a more capable and more expensive teacher model.

\paragraph{Ranking the best systems.} All methods that we compared order the best runs less reliably than the full set of runs. Over all runs, the ranking agreement of \method{} is at a level usually considered acceptable. On the ten runs that NIST ranks highest, $\tau$ is lower and varies by year for every automatic method. Appendix~\ref{app:more} lists these values. The labels are therefore suited to comparing systems of different quality. More work is needed to make \method{} and other automatic evaluators reliable for ranking high-performing retrieval systems.

\paragraph{Evaluation gaming and readability.} Readable programs can be read by system builders as well as by evaluators. A system builder who reads a released \method{} program could tune a ranker to its rules. We did not test this risk. We also did not measure how well people can read and check the programs. While by examination, the programs appear well structured, we did not perform usability testing of the generated code, which may be a promising extension.

\paragraph{Writer and LLM judge models.} Our results were generated using one family of models and on a strong program writer. All LLM judges and all but one program writer were closed frontier models from Anthropic Claude family. With an open-weight writer, label agreement dropped (Section~\ref{sec:analysis}). The results may therefore differ for other model families. A natural extension is to test open models optimized for code generation. Such models could reduce the reliance on frontier models for this task.

\section{Conclusions}
This paper presented \method{}, a method that replaces the per-document calls of an LLM judge with executable programs that an LLM writes once for each topic. The programs then label document pools of any depth without further LLM calls, and people can inspect and rerun them. We demonstrated that \method{} agrees with NIST assessors at the level of a frontier LLM judge, Claude Sonnet 5.5, on TREC DL 2019--2021.
When a few human labels per topic are available, \methodL{} agrees with NIST more than per-topic classifiers~\cite{gienapp2026topic} given the same labels on all seven collections, and its agreement is either higher than that of LLM judges given the same labels or within about 0.05 of it in Krippendorff's $\alpha$. The resulting labels rank the official TREC DL runs close to the ranking under NIST labels, with Kendall's $\tau$ of 0.87 to 0.94. Writing the programs with one of the most capable frontier models incurs a one-time cost of less than \$0.50 per topic, and does not depend on the number of result documents to label. 

Possible future directions include testing and improving the robustness of \method{} to adversarial content such as keyword stuffing, and writing programs that judge several topics, or topics from different domains to reduce program generation costs, without loss of label agreement. Another direction is to extend the method to long-form generative IR, where \method{} could provide low-cost rewards for training generators, and to web-scale labeling workloads. 

We report all the prompts, settings, and experimental details used in this study, and plan to publicly release all code and data.

\bibliographystyle{ACM-Reference-Format}
\bibliography{references_arxiv_edit6}

\appendix
\renewcommand{\topfraction}{0.95}\renewcommand{\bottomfraction}{0.9}\renewcommand{\textfraction}{0.05}\renewcommand{\floatpagefraction}{0.5}\renewcommand{\dbltopfraction}{0.95}\renewcommand{\dblfloatpagefraction}{0.5}\setcounter{topnumber}{4}\setcounter{dbltopnumber}{3}\setcounter{totalnumber}{6}

\section{Additional results}
\label{app:more}

\begin{table*}[t]
\caption{Binary $\alpha$ at grade $\geq$ 1 for the methods of Table~\ref{tab:main}. The conventions are those of Table~\ref{tab:main}.}
\label{tab:main-bin}
\small
\setlength{\tabcolsep}{4pt}
\begin{adjustbox}{max width=\textwidth}
\begin{tabular}{@{}lccccccc@{}}
\toprule
Method & DL 2019 & DL 2020 & DL 2021 & DL 2022 & DL 2023 & Robust04 & TREC-COVID \\
\midrule
\multicolumn{8}{@{}l}{\emph{No labels}} \\
\method{} & .565 & .442 & .582 & .507 & .435$^\dagger$ & .390 & .481 \\
Sonnet, listwise & .461$^\dagger$ & .443 & .556 & .463$^\dagger$ & .439$^\dagger$ & .466$^\dagger$ & .626$^\dagger$ \\
One-shot (1) & .470 & .407 & .413$^{**}$ & .435$^{*}$ & -- & $-$.547$^{**}$ & .267$^{**}$ \\
\midrule
\multicolumn{8}{@{}l}{\emph{16 labels}} \\
\methodL{} & .704 & .619 & .705 & .676 & .577$^\dagger$ & .420 & .575 \\
Sonnet, listwise & .525$^\dagger$ & .478$^{**}$ & .646$^{*}$ & .524$^\dagger$ & .518$^\dagger$ & .458$^\dagger$ & .620$^\dagger$ \\
Classifier & .636 & .506$^{**}$ & .596$^{**}$ & .541$^{**}$ & .452 & .123$^{**}$ & .432$^{**}$ \\
One-shot (16) & .509$^{**}$ & .404$^{**}$ & .500$^{**}$ & .515$^{**}$ & -- & $-$.074$^{**}$ & .349$^{**}$ \\
\midrule
\multicolumn{8}{@{}l}{\emph{64 labels}} \\
\methodL{} & .747$^\dagger$ & .702$^\dagger$ & .785$^\dagger$ & .718$^\dagger$ & .629$^\dagger$ & .498$^\dagger$ & .561$^\dagger$ \\
Sonnet, listwise & .627$^\dagger$ & .584$^\dagger$ & .698$^\dagger$ & .603$^\dagger$ & .603$^\dagger$ & .477$^\dagger$ & .624$^\dagger$ \\
\bottomrule
\end{tabular}
\end{adjustbox}
\end{table*}

\begin{table*}[t]
\caption{API cost per topic in USD at standard prices, including retries and failed calls, and mean time. \method{}: writing and running the programs. Sonnet: listwise judging of the test passages.}
\label{tab:cost}
\small
\setlength{\tabcolsep}{5pt}
\begin{adjustbox}{max width=\textwidth}
\begin{tabular}{@{}lcccccc@{}}
\toprule
 & \multicolumn{3}{c}{\method{}} & \multicolumn{3}{c}{Sonnet, listwise} \\
\cmidrule(lr){2-4}\cmidrule(l){5-7}
Collection & 0 labels & 16 labels & 64 labels & 0 labels & 16 labels & 64 labels \\
\midrule
DL 2019 & .177 & .210 & .320 & .066 & .071 & .150 \\
DL 2020 & .173 & .204 & .289 & .082 & .083 & .174 \\
DL 2021 & .173 & .217 & .342 & .053 & .059 & .124 \\
DL 2022 & .176 & .222 & .303 & .092 & .101 & .209 \\
DL 2023 & .184 & .232 & .553 & .085 & .098 & .213 \\
Robust04 & .231 & .306 & .669 & .324 & .460 & 1.038 \\
TREC-COVID & .225 & .489 & 1.011 & .233 & .344 & .767 \\
\midrule
Writing or judging (s/topic) & 60.0 & 77.2 & 88.0 & 2.5 & 3.5 & 4.0 \\
Grading (s/passage) & 0.002 & 0.003 & 0.004 & -- & -- & -- \\
\bottomrule
\end{tabular}
\end{adjustbox}
\end{table*}

\begin{table}[t]
\caption{Programs per collection: median [10th, 90th percentile]. Toolkit calls are calls of \texttt{jk.*}. Listed terms are strings in literal lists of three or more terms. We count all topics with programs. (16): \methodL{} with 16 labels.}
\label{tab:progstats}
\footnotesize
\setlength{\tabcolsep}{1.5pt}
\begin{adjustbox}{max width=\columnwidth}
\begin{tabular}{@{}lccccc@{}}
\toprule
Collection & Count & Lines & Regex calls & Toolkit calls & Listed terms \\
\midrule
DL 2019 & 120 & 74 [55, 92] & 6 [4, 10] & 2 [2, 3] & 14 [0, 71] \\
DL 2020 & 162 & 70 [54, 86] & 7 [5, 11] & 2 [2, 3] & 12 [0, 51] \\
DL 2021 & 150 & 72 [54, 90] & 7 [4, 9] & 3 [2, 4] & 12 [0, 60] \\
DL 2022 & 219 & 75 [56, 94] & 7 [4, 11] & 2 [2, 3] & 12 [0, 61] \\
Robust04 & 237 & 73 [51, 96] & 6 [3, 9] & 1 [1, 3] & 4 [0, 79] \\
TREC-COVID & 131 & 70 [50, 92] & 6 [1, 9] & 1 [1, 3] & 16 [0, 118] \\
\midrule
DL 19--21 (16) & 432 & 66 [50, 83] & 8 [4, 12] & 2 [1, 3] & 0 [0, 45] \\
\bottomrule
\end{tabular}
\end{adjustbox}
\end{table}

\begin{table}[t]
\caption{Variants and further baselines on DL 2019--2021 (residual passages): ordinal $\alpha$, and binary $\kappa$ at grade $\geq$ 2 on DL 2021. (n): n labels per topic. Stars: against \method{}, or \methodL{} with the same labels for labeled rows.}
\label{tab:pilot}
\small
\setlength{\tabcolsep}{1.6pt}
\begin{adjustbox}{max width=\columnwidth}
\begin{tabular}{@{}lcccc@{}}
\toprule
Method & DL19 & DL20 & DL21 & $\kappa$ DL21 \\
\midrule
\multicolumn{5}{@{}l}{\emph{No labels of the evaluated topic}} \\
\method{} & .613 & .483 & .653 & .527 \\
\method{}-1P & .589$^{**}$ & .442 & .632 & .519 \\
\method{}-R & .581 & .470 & .588$^{**}$ & .432 \\
\method{}-GR & .225$^{**}$ & .156$^{**}$ & .151$^{**}$ & .057 \\
\method{}-Global & .122$^{**}$ & .076$^{**}$ & $-$.091$^{**}$ & $-$.026 \\
\method{} (GPT-OSS) & .486$^{**}$ & .393$^{**}$ & .344$^{**}$ & .230 \\
\midrule
\multicolumn{5}{@{}l}{\emph{With labels of the evaluated topic}} \\
\methodL{} (4) & .694$^{**}$ & .587$^{**}$ & .708$^{**}$ & .548 \\
\methodL{} (8) & .704$^{**}$ & .618$^{**}$ & .729$^{**}$ & .559 \\
\methodL{} (12) & .734$^{**}$ & .625$^{**}$ & .747$^{**}$ & .581 \\
\methodL{} (16) & .756$^{**}$ & .649$^{**}$ & .750$^{**}$ & .588 \\
\methodL{} (3 BM25) & .612 & .514 & .663 & .500 \\
\methodL{} (16, GPT-OSS) & .521$^{**}$ & .467$^{**}$ & .438$^{**}$ & .223 \\
Classifier (4) & .582$^{**}$ & .465$^{**}$ & .513$^{**}$ & .387 \\
Classifier (3 BM25) & .431$^{**}$ & .330$^{**}$ & .511$^{**}$ & .429 \\
\midrule
\multicolumn{5}{@{}l}{\emph{LLM judges}} \\
Sonnet, pointwise & .581 & .524 & .617 & .506 \\
Sonnet, pointwise (16) & -- & -- & .706$^{*}$ & .584 \\
Haiku, pointwise & .504$^{**}$ & .447 & .580$^{**}$ & .486 \\
Haiku, pointwise (16) & .573$^{**}$ & .513$^{**}$ & .687$^{**}$ & .527 \\
\midrule
\multicolumn{5}{@{}l}{\emph{No LLM}} \\
Embedding cutoffs & .378$^{**}$ & .250$^{**}$ & .210$^{**}$ & .175 \\
Lexical classifier & .307$^{**}$ & .281$^{**}$ & .183$^{**}$ & .023 \\
BM25 cutoffs & .151$^{**}$ & .056$^{**}$ & $-$.048$^{**}$ & $-$.006 \\
\bottomrule
\end{tabular}
\end{adjustbox}
\end{table}

\begin{table}[t]
\caption{System rankings on DL 2019--2021 (official runs, nDCG@10). \method{} + NIST: \method{} labels with the 16 NIST labels inserted. Stars: against the first row of each block; GPT-OSS rows against the same method with Opus 5.5.}
\label{tab:tau-dl}
\footnotesize
\setlength{\tabcolsep}{1.1pt}
\begin{adjustbox}{max width=\columnwidth}
\begin{tabular}{@{}lcccccc@{}}
\toprule
 & \multicolumn{2}{c}{DL19 (36 runs)} & \multicolumn{2}{c}{DL20 (59 runs)} & \multicolumn{2}{c}{DL21 (62 runs)} \\
\cmidrule(lr){2-3}\cmidrule(lr){4-5}\cmidrule(l){6-7}
Labels & $\tau$ & $\tau_{AP}$ & $\tau$ & $\tau_{AP}$ & $\tau$ & $\tau_{AP}$ \\
\midrule
\method{} & .905 & .865 & .903 & .793 & .869 & .866 \\
\method{}-1P & .927 & .859 & .889 & .772 & .891 & .886 \\
\method{}-R & .892 & .822 & .917 & .837 & .890 & .874 \\
\midrule
Sonnet, pointwise & .917 & .909 & .952$^{*}$ & .880 & .896 & .884 \\
Sonnet, listwise & -- & -- & .940 & .855 & .901 & .884 \\
Haiku, pointwise & .927 & .886 & .917 & .758 & .910 & .867 \\
Embedding cutoffs & .867 & .816 & .856 & .701 & .702$^{**}$ & .564$^{**}$ \\
Lexical classifier & .027$^{**}$ & .090$^{**}$ & .302$^{**}$ & .301$^{**}$ & .436$^{**}$ & .286$^{**}$ \\
BM25 cutoffs & $-$.559$^{**}$ & $-$.354$^{**}$ & $-$.174$^{**}$ & $-$.225$^{**}$ & .171$^{**}$ & .193$^{**}$ \\
\midrule
\methodL{} (16) & .927 & .854 & .944 & .914 & .926 & .916 \\
\method{} + NIST (16) & .902 & .895 & .911 & .838 & .889 & .886 \\
Classifier (16) & .917 & .799 & .859$^{*}$ & .676$^{*}$ & .854$^{*}$ & .747$^{*}$ \\
One-shot (16) & .913 & .847 & .917 & .775$^{*}$ & .881 & .810 \\
Sonnet, listwise (16) & -- & -- & .940 & .824 & .916 & .905 \\
Sonnet, pointwise (16) & -- & -- & -- & -- & .899 & .847 \\
Haiku, pointwise (16) & .924 & .912 & .932 & .819 & .926 & .922 \\
\midrule
\methodL{} (64 BM25) & .975 & .973 & .973 & .951 & .912 & .901 \\
Classifier (64 BM25) & .962 & .931 & .942 & .852 & .911 & .840 \\
\midrule
\method{} (GPT-OSS) & .543$^{**}$ & .305$^{*}$ & .833$^{*}$ & .712 & .856 & .824 \\
\methodL{} (16, GPT-OSS) & .898 & .804 & .933 & .911 & .862$^{*}$ & .793 \\
\bottomrule
\end{tabular}
\end{adjustbox}
\end{table}

\begin{table}[t]
\caption{Pool reuse on DL 2019--2021, one run group left out at a time: mean $|\Delta|$ nDCG@10 of its runs ($\times 100$) and $\tau$ with the NIST ranking. Stars: group-bootstrap interval against \methodL{} (16).}
\label{tab:holes}
\footnotesize
\setlength{\tabcolsep}{2.2pt}
\begin{adjustbox}{max width=\columnwidth}
\begin{tabular}{@{}lcccccc@{}}
\toprule
 & \multicolumn{2}{c}{DL19 (11 groups)} & \multicolumn{2}{c}{DL20 (19)} & \multicolumn{2}{c}{DL21 (15)} \\
\cmidrule(lr){2-3}\cmidrule(lr){4-5}\cmidrule(l){6-7}
Removed passages & $|\Delta|$ & $\tau$ & $|\Delta|$ & $\tau$ & $|\Delta|$ & $\tau$ \\
\midrule
Non-relevant & 2.16$^{*}$ & 0.879 & 1.19$^{*}$ & 0.978 & 3.98$^{*}$ & 0.863 \\
\method{} & 0.81$^{*}$ & 0.908 & 0.88$^{*}$ & 0.986 & 0.86$^{*}$ & 0.947 \\
\methodL{} (16) & 0.25 & 0.971 & 0.25 & 0.987 & 0.46 & 0.967 \\
Sonnet, pointwise & 0.69$^{*}$ & 0.940 & 0.80$^{*}$ & 0.981 & 1.73$^{*}$ & 0.899 \\
Classifier (16) & 1.12$^{*}$ & 0.943 & 1.00$^{*}$ & 0.952 & 1.07$^{*}$ & 0.951 \\
One-shot (16) & 1.28$^{*}$ & 0.956 & 1.00$^{*}$ & 0.977 & 0.80$^{*}$ & 0.952 \\
\bottomrule
\end{tabular}
\end{adjustbox}
\end{table}

\begin{table*}[t]
\caption{Differences on the residual passages with 95\% topic-bootstrap intervals: ordinal $\alpha$ (top) and $\tau$ of the run rankings (bottom).}
\label{tab:diffs}
\footnotesize
\setlength{\tabcolsep}{5pt}
\begin{adjustbox}{max width=\textwidth}
\begin{tabular}{@{}lccc@{}}
\toprule
 & DL19 & DL20 & DL21 \\
\midrule
\multicolumn{4}{@{}l}{\emph{Ordinal $\alpha$: \method{} minus}} \\
Sonnet, pointwise & +.032 [$-$.028, +.092] & $-$.042 [$-$.090, +.006] & +.036 [$-$.020, +.100] \\
Sonnet, listwise & -- & $-$.014 [$-$.061, +.029] & +.018 [$-$.045, +.098] \\
Haiku, pointwise & +.109 [+.050, +.167] & +.036 [$-$.015, +.091] & +.073 [+.014, +.137] \\
\multicolumn{4}{@{}l}{\emph{Ordinal $\alpha$: row minus \method{}}} \\
\methodL{} (4) & +.081 [+.028, +.160] & +.104 [+.049, +.165] & +.054 [+.024, +.089] \\
\methodL{} (16) & +.143 [+.069, +.240] & +.166 [+.098, +.245] & +.096 [+.056, +.139] \\
\multicolumn{4}{@{}l}{\emph{Ordinal $\alpha$: \methodL{} (16) minus, same labels}} \\
Classifier & +.050 [+.008, +.096] & +.106 [+.049, +.166] & +.084 [+.037, +.140] \\
One-shot (16) & +.179 [+.130, +.233] & +.190 [+.124, +.272] & +.156 [+.095, +.230] \\
Sonnet, listwise & -- & +.117 [+.062, +.181] & +.037 [+.003, +.075] \\
Haiku, pointwise & +.183 [+.117, +.262] & +.135 [+.081, +.195] & +.062 [+.023, +.115] \\
\midrule
\multicolumn{4}{@{}l}{\emph{$\tau$: \method{} minus}} \\
Sonnet, pointwise & $-$.013 [$-$.127, +.051] & $-$.049 [$-$.127, $-$.003] & $-$.028 [$-$.103, +.031] \\
Haiku, pointwise & $-$.022 [$-$.130, +.051] & $-$.014 [$-$.136, +.026] & $-$.041 [$-$.108, +.021] \\
\multicolumn{4}{@{}l}{\emph{$\tau$: \methodL{} (16) minus}} \\
\method{} & +.022 [$-$.035, +.133] & +.041 [+.007, +.151] & +.057 [$-$.001, +.106] \\
\method{} + NIST & +.025 [$-$.048, +.095] & +.033 [$-$.006, +.117] & +.037 [$-$.006, +.083] \\
Classifier (16) & +.010 [$-$.066, +.090] & +.085 [+.015, +.131] & +.072 [+.008, +.114] \\
\bottomrule
\end{tabular}
\end{adjustbox}
\end{table*}

\begin{table}[t]
\caption{The DL 2020 setting of Gienapp et al.: binary $\alpha$ on their evaluation passages, without our labeled passages (5,950) and all (6,575). $^\dagger$Recomputed from their released predictions. $^\ddagger$64 labels from outside their evaluation set (Section~\ref{sec:analysis}).}
\label{tab:gienapp}
\small
\setlength{\tabcolsep}{3pt}
\begin{adjustbox}{max width=\columnwidth}
\begin{tabular}{@{}lccc@{}}
\toprule
Method & Labels & $\alpha$ (5,950) & $\alpha$ (6,575) \\
\midrule
Topic classifier$^\dagger$ & 64 & .811 & .810 \\
Topic classifier$^\dagger$ & 128 & .812 & .808 \\
Topic classifier$^\dagger$ & 192 & .834 & .831 \\
Topic classifier$^\dagger$ & 256 & \textbf{.877} & \textbf{.877} \\
\midrule
\methodL{} & 4 & .623 & -- \\
\methodL{} & 8 & .658 & -- \\
\methodL{} & 12 & .665 & -- \\
\methodL{} & 16 & .674 & -- \\
\methodL{} (3 BM25) & 3 & .511 & -- \\
\method{} & 0 & .485 & .496 \\
\midrule
GPT-4o, UMBRELA$^\dagger$ & 0 & .526 & .539 \\
Qwen3-Next, 8-shot$^\dagger$ & 0 & .566 & .575 \\
Sonnet 5.5, pointwise & 0 & .557 & .571 \\
Haiku 5.5, pointwise & 0 & .477 & .496 \\
Haiku 5.5, pointwise (16) & 16 & .532 & -- \\
Sonnet 5.5, listwise & 0 & .515 & -- \\
Sonnet 5.5, listwise (16) & 16 & .540 & -- \\
\midrule
\methodL{} (64, their pool)$^\ddagger$ & 64 & -- & .330 \\
\quad with label-mix note$^\ddagger$ & 64 & -- & .407 \\
\bottomrule
\end{tabular}
\end{adjustbox}
\end{table}

\begin{table}[t]
\caption{Error types (\%): exact agreement, two or more grades off, NIST 3 labeled 0, NIST 0 labeled 2 or 3, and the mean of label minus NIST grade. DL 2019--2021: residual passages; DL 2022: the 7,402 passages the Sonnet judge labeled.}
\label{tab:errors}
\footnotesize
\setlength{\tabcolsep}{2.4pt}
\begin{adjustbox}{max width=\columnwidth}
\begin{tabular}{@{}lccccc@{}}
\toprule
Labels & Exact & $\geq$2 off & 3$\to$0 & 0$\to\geq$2 & Mean diff. \\
\midrule
\multicolumn{6}{@{}l}{\emph{DL 2019}} \\
\method{} & 61.9 & 8.4 & 26.4 & 4.0 & +0.01 \\
\methodL{} (16) & 69.8 & 4.6 & 1.2 & 2.5 & $-$0.02 \\
Sonnet, pointwise & 56.8 & 7.2 & 24.0 & 4.1 & +0.11 \\
Haiku, pointwise & 53.6 & 8.6 & 23.6 & 6.1 & +0.17 \\
Classifier (16) & 64.7 & 6.7 & 0.8 & 7.0 & +0.14 \\
One-shot (16) & 57.2 & 7.9 & 7.7 & 6.3 & +0.06 \\
\midrule
\multicolumn{6}{@{}l}{\emph{DL 2020}} \\
\method{} & 65.0 & 8.5 & 28.3 & 5.0 & +0.11 \\
\methodL{} (16) & 73.3 & 5.3 & 8.1 & 3.6 & +0.04 \\
Sonnet, pointwise & 64.3 & 6.8 & 17.0 & 4.7 & +0.19 \\
Haiku, pointwise & 57.9 & 7.5 & 10.4 & 6.5 & +0.25 \\
Classifier (16) & 67.0 & 9.1 & 6.2 & 9.7 & +0.22 \\
One-shot (16) & 61.5 & 8.7 & 1.7 & 6.2 & +0.16 \\
\midrule
\multicolumn{6}{@{}l}{\emph{DL 2021}} \\
\method{} & 54.9 & 8.0 & 2.4 & 7.8 & +0.12 \\
\methodL{} (16) & 63.8 & 3.9 & 1.3 & 4.5 & +0.03 \\
Sonnet, pointwise & 51.1 & 7.8 & 0.1 & 9.1 & +0.34 \\
Haiku, pointwise & 49.0 & 8.1 & 0.0 & 12.4 & +0.34 \\
Classifier (16) & 57.2 & 8.0 & 0.3 & 11.6 & +0.22 \\
One-shot (16) & 54.8 & 8.7 & 2.6 & 9.1 & +0.04 \\
\midrule
\multicolumn{6}{@{}l}{\emph{DL 2022}} \\
\method{} & 53.0 & 7.7 & 6.5 & 6.3 & +0.13 \\
\methodL{} (16) & 63.0 & 5.6 & 2.7 & 5.6 & +0.09 \\
Sonnet, pointwise & 55.1 & 5.3 & 1.9 & 4.9 & +0.27 \\
Classifier (16) & 45.0 & 17.0 & 0.8 & 25.2 & +0.58 \\
\bottomrule
\end{tabular}
\end{adjustbox}
\end{table}

\begin{table}[t]
\caption{Label sets and classifier variants, ordinal $\alpha$. DL 2019--2021: residual passages without the 64 BM25 passages. Oracle: per-topic cut points that reproduce the NIST grade distribution. Stars: against \methodL{} (16); (64 BM25) rows on the held-out collections are evaluated without their 64 labels (TREC-COVID: 42 topics) and not tested.}
\label{tab:clsvar}
\footnotesize
\setlength{\tabcolsep}{1.4pt}
\begin{adjustbox}{max width=\columnwidth}
\begin{tabular}{@{}lcccccc@{}}
\toprule
 & DL19 & DL20 & DL21 & DL22 & Rob04 & COVID \\
\midrule
\method{} & .616$^{**}$ & .463$^{**}$ & .665$^{**}$ & -- & -- & -- \\
\methodL{} (16) & .767 & .643 & .762 & .702 & .424 & .626 \\
\methodL{} (64 BM25) & .693 & .569 & .715 & -- & -- & -- \\
\midrule
Classifier (16) & .733 & .539$^{*}$ & .688$^{*}$ & .573$^{**}$ & .131$^{**}$ & .506$^{**}$ \\
\quad oracle cut points & -- & -- & -- & .673$^{*}$ & .385$^{**}$ & .562$^{**}$ \\
\quad weighted & -- & -- & -- & .575$^{**}$ & .131$^{**}$ & .510$^{**}$ \\
\quad wtd., oracle cuts & -- & -- & -- & .669$^{*}$ & .386$^{**}$ & .566$^{**}$ \\
\quad MaxP & -- & -- & -- & -- & $-$.118$^{**}$ & -- \\
\quad MaxP, oracle cuts & -- & -- & -- & -- & .404$^{**}$ & -- \\
\quad MaxP, wtd., oracle & -- & -- & -- & -- & .408$^{**}$ & -- \\
Classifier (4) & -- & -- & -- & .474$^{**}$ & .054$^{**}$ & .488$^{**}$ \\
Classifier (16 BM25) & -- & -- & -- & .426$^{**}$ & .095$^{**}$ & .366$^{**}$ \\
Classifier (64 BM25) & .639 & .534$^{*}$ & .702$^{*}$ & .564 & .203 & .494 \\
\midrule
One-shot (16) & -- & -- & -- & .532$^{**}$ & $-$.063$^{**}$ & .393$^{**}$ \\
\quad oracle cut points & -- & -- & -- & .587$^{**}$ & .254$^{**}$ & .474$^{**}$ \\
Sonnet, pointwise & .555$^{**}$ & .504$^{**}$ & .605$^{**}$ & -- & -- & -- \\
\bottomrule
\end{tabular}
\end{adjustbox}
\end{table}

\paragraph{Per-collection values for the main comparisons.} This paragraph gives the per-collection values behind Section~\ref{sec:main}. Without labels, \method{} is below the listwise Sonnet judge on DL 2023 by 0.048, on Robust04 by 0.090 and on TREC-COVID by 0.171. With 16 labels, \methodL{} is below this judge on Robust04 by 0.049 and on TREC-COVID by 0.052. With 64 labels, \methodL{} is above the judge by 0.059 to 0.087 on DL 2019--2022. It is within 0.04 of the judge on DL 2023 and Robust04. It is below the judge on TREC-COVID by 0.073. On DL 2019--2021, \method{} is above the pointwise Haiku judge by 0.109 on DL 2019 and by 0.073 on DL 2021. The intervals of both differences exclude 0. One-shot (1) is 0.124 below \method{} on DL 2019 and 0.163 below on DL 2021. On the documents that the pointwise Sonnet judge labeled, the judge reaches 0.577, 0.500 and 0.693 on DL 2022, Robust04 and TREC-COVID. \method{} reaches 0.513, 0.367 and 0.496. \methodL{} reaches 0.628, 0.442 and 0.661. On DL 2022, these documents are the 7,402 judged passages in the top 10 of some run. On Robust04 and TREC-COVID, they are 50 random documents per topic.

\paragraph{Per-collection values for topic labels.} This paragraph gives the per-collection values behind Section~\ref{sec:labels}. Prompting \methodL{} with just four labels per topic, one per grade, raises its $\alpha$ by 0.081, 0.104 and 0.054 on DL 2019--2021. Sixteen labels raise $\alpha$ by 0.143, 0.166 and 0.096. Going from 12 to 16 labels adds only 0.003 to 0.024. The mean difference from the NIST grade is +0.01 to +0.12 for \method{}. With 16 labels, this difference falls to $-$0.02 to +0.04. The three BM25 top passages change $\alpha$ by only $-$0.001 to +0.031. The labels raise the $\alpha$ of the judges by 0.035 to 0.107. \methodL{} remains above the pointwise Haiku judge by 0.062 to 0.183. It remains above the pointwise Sonnet judge by 0.043 on DL 2021. With labels, the mean difference from the NIST grade on DL 2021 is +0.25 for the pointwise Sonnet judge and +0.16 for the listwise Sonnet judge. The mean difference for \methodL{} is +0.03.

\paragraph{Per-collection values for the supervised baselines.} This paragraph gives the per-collection values behind Section~\ref{sec:cls}. On DL 2019--2021, the gap in ordinal $\alpha$ between \methodL{} and the classifier is 0.050, 0.106 and 0.084. The gap in binary $\alpha$ is 0.067, 0.113 and 0.110. With 4 labels, the gap in ordinal $\alpha$ grows to 0.112, 0.122 and 0.195. The mean difference of the classifier from the NIST grade is +0.14 to +0.22. One-shot (16) is 0.156 to 0.190 below \methodL{}. With 64 BM25 labels, the classifier reaches 0.639, 0.534 and 0.702. \methodL{} with the same 64 BM25 labels is 0.012 to 0.054 above the classifier. These differences are not significant. In the published setting of Gienapp et al., the binary $\alpha$ of their classifiers is 0.81 to 0.88. \methodL{} reaches 0.674. On DL 2022, Robust04 and TREC-COVID, \methodL{} is above the classifier on 45 of 65, 69 of 79 and 32 of 36 topics. It is above One-shot (16) on 48, 75 and 36 of these topics. On average, 9 of the 16 stratified labels of a Robust04 topic are relevant. The classifier reaches at most 0.203 on Robust04 with any label set. With oracle cut points, the classifier reaches 0.673, 0.385 and 0.562. One-shot (16) reaches 0.587, 0.254 and 0.474.

\paragraph{Per-collection values for system rankings.} This paragraph gives the per-collection values behind Section~\ref{sec:rank}. Without labels, \method{} reaches $\tau$ of 0.905, 0.903, 0.869 and 0.882 on DL 2019--2022. These values are within the intervals of the pointwise Sonnet and Haiku judges on all years except DL 2020. On DL 2020, the Sonnet judge is 0.049 higher. With 16 labels, \methodL{} reaches 0.927, 0.944, 0.926 and 0.925. As a control, we inserted the same 16 NIST grades into the \method{} labels. This control reaches 0.902, 0.911 and 0.889 on DL 2019--2021. \methodL{} is 0.025 to 0.037 above the control. The intervals of these differences include 0. With the same labels, the classifier reaches 0.917, 0.859, 0.854 and 0.866. The average increase of nDCG@10 under the classifier labels is 0.14 on DL 2021 and 0.20 on DL 2022. The embedding cutoffs have $\alpha$ of only 0.378 and 0.250 on DL 2019 and 2020. Their $\tau$ is still 0.867 and 0.856. In the reuse experiment, \methodL{} labels change the nDCG@10 of the left-out runs by 0.0025 to 0.0046 on average. \method{} labels change nDCG@10 by 0.0081 to 0.0088. Counting the passages as non-relevant changes it by 0.0119 to 0.0398.

\paragraph{Ranking the best runs.} On the ten runs that NIST ranks highest, ranking agreement is much lower than over all runs. On DL 2022, $\tau$ on these runs is 0.244 for \method{}. With 16 labels, it is 0.689 for \methodL{}, 0.511 for the topic classifier and 0.600 for One-shot. On DL 2019, 2020 and 2021, $\tau$ on these runs is 0.733, 0.467 and 0.857 for \method{}.

\paragraph{Cost, programs and errors.} This paragraph gives the values behind Sections~\ref{sec:cost} and~\ref{sec:analysis}. Writing the programs costs \$0.17 to \$0.24 per topic without labels. With 16 labels, writing costs \$0.20 to \$0.49 per topic. The listwise Sonnet judge with 16 labels costs \$0.53 to \$2.31 per 1,000 judgments. Figure~\ref{fig:cost} uses \$0.60 for TREC DL and \$2.00 for Robust04 and TREC-COVID. The break-even points are about 350 and 200 documents per topic. In Figure~\ref{fig:labels}, 0 labels is \method{}, the values are on the residual passages, and the judge markers are shifted sideways. The median program has 66 to 75 lines and two functions. The median program contains 6 to 8 regular expressions and one to three toolkit calls. On Robust04, 92\% of the programs look for one sentence that meets the conditions of the topic. With 16 labels, the programs list fewer terms and use one more regular expression. With 16 labels each, \methodL{} is ahead of the classifier by more than 0.02 on 84 of 142 DL 2019--2021 topics. It is behind on 42 topics. \methodL{} gains 0.6 to 1.0 on the named-entity topics. All Spearman correlations of the per-topic difference with pool size, the share of relevant passages and query length satisfy $|\rho| \leq 0.15$ with $p>0.07$. In the larger set of programs that contains the prompt variants of Appendix~\ref{app:variants}, going from three to five programs adds only 0.005 to 0.009. Going from five to seven programs adds 0.002 to 0.007. Two rewrites of the same program differ by up to 0.024 on one collection. Each writer call costs about one third of the writing cost. With GPT-OSS-120B as the writer, \method{} reaches 0.486, 0.393 and 0.344 on DL 2019--2021. \methodL{} with 16 labels reaches 0.521, 0.467 and 0.438. The mean difference of the GPT-OSS programs from the NIST grade is $-$0.14 to $-$0.48. On DL 2019 and 2020, \method{} gives grade 0 to 26\% and 28\% of the passages that NIST graded 3. The Sonnet judge gives grade 0 to 24\% and 17\% of these passages. Four topics per year hold 90 to 95\% of these errors. With labels, the share falls to 1.2\%, 8.1\% and 1.3\% on DL 2019--2021. On DL 2019--2021, \methodL{} gives grade 2 or 3 to 2.5 to 4.5\% of the passages that NIST graded 0. The classifier gives these grades to 7.0 to 11.6\% of these passages. On the 7,402 DL 2022 passages that the pointwise Sonnet judge also labeled, \method{}, \methodL{} and the Sonnet judge are two or more grades off on 7.7\%, 5.6\% and 5.3\%. The classifier gives grades of 2 or 3 to 25.2\% of the NIST grade-0 passages. The other three methods give them to at most 6.3\%. \method{} gives grade 0 to 10.6\% of the passages that NIST graded 2 or 3. The other three methods give grade 0 to at most 5.8\% of these passages.

\paragraph{Classifier with the released settings.} With the released training settings, the classifier labels almost every passage relevant. Section~\ref{sec:setup} describes these settings. We ran the classifier with these settings on matched test passages in the separate runs. With 16 or 64 labels, its binary $\alpha$ is $-$0.62 to $-$0.13 on six of the seven collections of Table~\ref{tab:main}. On TREC-COVID, its binary $\alpha$ is 0.04 with 16 labels and 0.09 with 64 labels.

\paragraph{The published setting of Gienapp et al.} In this setting, the classifiers of Gienapp et al. agree with NIST more than every other method. We use their DL 2020 evaluation passages and their released predictions. Table~\ref{tab:gienapp} lists the results. Their classifiers trained on 64 to 256 labels per topic reach $\alpha$ of 0.81 to 0.88. \methodL{} with 16 stratified labels reaches 0.674. \method{} reaches 0.485. GPT-4o with the UMBRELA prompt reaches 0.526. The Sonnet judge reaches 0.557. The share of relevant passages in this evaluation set ranges from 0\% to 100\% across topics. Pooled $\alpha$ is therefore higher for a method that labels the correct share in each topic. On the 28 topics with more than 50 relevant passages, the 64-label classifier reaches only 0.681. The training labels of these classifiers are 94\% non-relevant. With 64 of these labels, the programs label too few passages relevant. Appendix~\ref{app:details} describes this experiment.

\paragraph{What did not help.} We tried four ways to correct program errors after we fixed the main method. None of them raised agreement on passages outside those used for the correction. First, the writer revised the shared code that turns the answers into a grade. We split DL 2019 and 2020 into 57 training topics and 37 held-out topics for this test. In four rounds, held-out $\alpha$ changed by $-$0.014 to +0.003. Second, the writer revised a block of guidance in its own prompt. The one round within the 350-word limit lowered held-out $\alpha$ by 0.010. Third, the writer revised each \methodL{} program after seeing its errors on the 16 labeled passages. The errors on these passages fell from 9--12\% to 3--4\%. On the other passages, $\alpha$ changed by only +0.002, $-$0.005 and +0.006. Fourth, the writer induced written guidelines per year from 10 topics with NIST labels. These guidelines lowered $\alpha$ on the other topics by 0.017 on DL 2019 and by 0.037 on DL 2021. The interval of the DL 2021 change excludes 0. We also fitted a rule per year that maps the four answers to grades. This rule lost 0.02 to 0.05 $\alpha$ on topics it was not fitted on. The released package will contain the guidelines and revision logs.

\paragraph{Errors of each method on DL 2022.}\label{app:errors} Most large errors are made by only one of the four methods. We count passages that are two or more grades off the NIST grade on the 7,402 passages of Table~\ref{tab:errex}. Only the classifier is off on 884 passages. Only \method{} is off on 226 passages. Only the Sonnet judge is off on 130 passages. Only \methodL{} is off on 91 passages. All four methods are off on 47 passages. The over-grading of the classifier occurs on 63 of the 65 topics. The missed grades of \method{} occur on 36 topics. One topic, \emph{example of what a family advocate does}, holds 31 of these 149 passages. With 16 labels, 37 of these passages receive grade 2 or 3. Each program error in Table~\ref{tab:errex} traces back to a term list or pattern in the code.

\section{Method details and further analyses}
\label{app:details}

\paragraph{Changes after the pilot.} DL 2021 served as the pilot test. After the pilot, we changed the program to return answers to the grade questions, replaced the single rubric program with the mean of three per-query programs, and added \methodL{} and all label-based baselines.

\paragraph{Classifier settings.} We tuned three settings of the topic classifier because the released settings fail on our labels. The classifier trains one LoRA adapter per topic on monoT5-base with rank 64 and $\alpha=128$. We keep the released optimizer settings and the mean squared error loss. With the released input pairs, 10 epochs and example weights of 0.95 and 0.05, the classifiers labeled almost every DL 2019 passage relevant. We tuned three settings on 8 DL 2019 topics. These settings are the input format of monoT5, 60 epochs and the inverse share of relevant labels as example weights. We report a graded version that trains on the grade divided by 3. It cuts the score at $(g-0.5)/3$ for $g=1,2,3$. On the tuning topics, the graded version reached binary $\alpha$ of 0.62 against 0.57 for the binary version. The graded version reads the first 512 tokens of each document. Table~\ref{tab:clsvar} reports both variants. On the 32 DL 2019 topics outside the tuning topics, \methodL{} is 0.047 above the classifier.

\paragraph{Rounding the mean of three programs.} The best rounding direction depends on the bias of single programs. On DL 2019--2021, single programs grade too high. On DL 2021, they are 0.27 too high on average. Rounding the mean down then gives higher $\alpha$ than rounding to the nearest grade. The gain is 0.003, 0.006 and 0.017. On the held-out collections, single programs are close to unbiased on DL 2022 and Robust04. They grade too low on TREC-COVID by 0.15 to 0.19. Rounding to the nearest grade then gives 0.557, 0.413 and 0.576. \method{} with rounding down reaches 0.549, 0.392 and 0.508. The difference is significant at $p<0.01$ on Robust04 and TREC-COVID. The rounding direction should follow the bias of the programs. Labeled passages can measure this bias when they exist.

\paragraph{Disagreement among the programs.} Disagreement among the programs of a topic does not identify the topics on which the programs fail. On the held-out collections, the programs of a topic disagree on 23.6\%, 6.7\% and 21.7\% of its evaluation documents on average. This share is not correlated with the $\alpha$ of the topic. The Spearman correlations are $-$0.05, +0.14 and +0.11 with $p>0.2$.

\paragraph{Held-out diagnostics.} The classifier needs 16 stratified labels to reach the level of \method{} on DL 2022 and TREC-COVID. With 16 labels, it reaches 0.573 and 0.506 against 0.549 and 0.507 for \method{}. One-shot (16) is level with \method{} on DL 2022 and below it on the other two collections. Its scores still rank the documents of a topic well. The mean per-topic AUC for NIST grade $\geq$ 1 is 0.74 to 0.78. MaxP scoring takes the maximum score over windows of 150 words. It labels 43\% of the Robust04 documents relevant. The mean of per-topic $\alpha$ orders the methods as pooled $\alpha$ does. On DL 2022 with one passage per distinct text, \methodL{} reaches 0.669 and the classifier 0.518.

\paragraph{Rubric programs.} Rubric programs agree with NIST no better than per-query programs. Induced rubrics improve LLM judges in prior work~\cite{farzi2025criteria}. In our setting, the writer compiles an induced rubric into lexical checks of a program. \method{}-R agrees with NIST as well as \method{}-1P on the development years. The differences are $-$0.008 and +0.028 and not significant. On DL 2021, \method{}-R is 0.044 below \method{}-1P. Each criterion becomes a lexical check that can fail. A failed check at the lowest level sets the grade to 0. On DL 2019 and 2020, \method{}-R gave grade 0 to 23--28\% of the passages NIST graded 3 in this way. The per-query program already decomposes relevance into the grade questions of the guidelines. An induced rubric repeats these questions and adds checks that can fail. This result concerns rubrics compiled into programs. We did not test rubrics applied by an LLM judge.

\paragraph{What the program should return.} A program should return the answers to the grade questions or a grade. A binary output lowers agreement. We compared the three outputs with the 64 labels of the DL 2020 setting of Table~\ref{tab:gienapp}. A program that returns a binary label is below the four answers of \method{} in binary $\alpha$ by 0.113 for a single program. The interval of this difference is 0.059 to 0.170. For the mean of three programs, the difference is 0.191. Its interval is 0.124 to 0.265. A program that returns a grade 0--3 does not differ from the four answers. The difference is $-$0.009 with an interval of $-$0.054 to +0.039.

\paragraph{Which passages to label.} The labels should cover all grades. In the DL 2020 setting, the 64 labels had to come from judged passages outside the evaluation set of the released classifiers. Only 4.7 of these 64 labels per topic were relevant on average. With these labels, the programs label 23--26\% of the passages relevant. NIST labels 50\% of them relevant. The programs reach binary $\alpha$ of 0.33 to 0.38. These values are below \method{} without labels at 0.50. We told the writer that relevant passages are under-represented among the labels. This note raised $\alpha$ to 0.41 to 0.46. The intervals of these gains include 0. The BM25 top-3 result of Section~\ref{sec:labels} points the same way.

\section{Reproduction details}
\label{app:release}

\paragraph{Prompt variants.}\label{app:variants} None of three prompt variants improved a single per-query program. We tested the variants after the pilot. The first variant uses stemmed phrase matching in any order with static checks. The second variant adds embedding similarity from bge-small-en-v1.5. The third variant uses one shared grading function with per-topic settings. On DL 2021, the variants reached $\alpha$ of 0.615, 0.606 and 0.581 on all judged passages of the pilot run. The program of Section~\ref{sec:method} reached 0.636. We use the variant programs only in the pooled analysis of the number of programs.

\paragraph{Collections and example topics.} Table~\ref{tab:collstats} gives the size of each collection and the grade distribution of its evaluation passages. The example topics are 1063750, 1103812 and 168216 for DL 2019 and DL 2020; 1040198, 1111577 and 1129560 for DL 2021; 2003976, 2004237 and 2044423 for DL 2022; 2004980, 2006402 and 3100119 for DL 2023; 612, 620 and 633 for Robust04; and 5, 11 and 42 for TREC-COVID. The collections have 40, 54 and 50 topics for DL 2019, 2020 and 2021. The test collections DL 2022, DL 2023, TREC-COVID and Robust04 have 73, 79, 47 and 79 topics after we remove the example topics. Table~\ref{tab:main} reports the real and synthetic topics of DL 2023 together. On 8 DL 2022 topics, NIST attached the same 42,393 short passages to the topic with one grade per topic. We evaluate DL 2022 on the other 65 topics. On 5 TREC-COVID topics about SARS-CoV-2 biology, the API refused some writer requests on every attempt. We therefore evaluate TREC-COVID on the other 42 topics. With 16 labels in the prompt, the writer reply was cut off for 6 more topics. The comparisons with 16 labels therefore use 36 topics. TREC-COVID documents consist of the title and abstract of each article. We cut DL 2022 and 2023 passages to 4,000 characters. We cut Robust04 documents to 8,000 characters.

\paragraph{Prompt and tools.} Figure~\ref{fig:prompt} gives the writer prompt for TREC DL with the grade definitions and grade distribution of DL 2019. From DL 2021 on, the grade definitions are the NIST guideline text. Figure~\ref{fig:policy} gives the tool policy. Table~\ref{tab:settings} lists the model settings, the sandbox limits and the seed.

\begin{table}[tp]
\caption{Collections: evaluated topics, judged passages of these topics, evaluation passages and their NIST grade distribution. Evaluation passages exclude the labeled passages (Section~\ref{sec:setup}). TREC-COVID comparisons with 16 labels use 36 of the 42 topics. DL 2023: the cells of Table~\ref{tab:main} come from separate runs on other passages.}
\label{tab:collstats}
\footnotesize
\setlength{\tabcolsep}{3pt}
\begin{adjustbox}{max width=\columnwidth}
\begin{tabular}{@{}lrrrc@{}}
\toprule
Collection & Topics & Judged & Evaluation & Grades 0/1/2/3 (\%) \\
\midrule
DL 2019 & 40 & 8,145 & 7,407 & 59 / 16 / 18 / 7 \\
DL 2020 & 54 & 11,386 & 10,397 & 71 / 16 / 8 / 5 \\
DL 2021 & 50 & 10,216 & 9,284 & 41 / 28 / 21 / 10 \\
DL 2022 & 65 & 33,627 & 31,595 & 68 / 20 / 9 / 4 \\
DL 2023 & 79 & 21,460 & 19,029 & 63 / 20 / 9 / 8 \\
Robust04 & 79 & 64,314 & 61,930 & 96.5 / 2.6 / 0.9 / -- \\
TREC-COVID & 42 & 58,522 & 57,186 & 62 / 16 / 22 / -- \\
\bottomrule
\end{tabular}
\end{adjustbox}
\end{table}

\begin{table}[tp]
\caption{Run settings. Model IDs are those of the Anthropic API.}
\label{tab:settings}
\footnotesize
\begin{adjustbox}{max width=\columnwidth}
\begin{tabular}{@{}p{0.25\columnwidth}p{0.71\columnwidth}@{}}
\toprule
Program writer & \texttt{claude-opus-5-5}, effort medium, thinking on (the model default), at most 32,000 output tokens including thinking \\
Writer inputs & 5 unlabeled pool passages of up to 1,200 characters; for \methodL{}, labeled passages of up to 600 characters on DL and 1,200 on Robust04 and TREC-COVID \\
Repair & one repair call when a program fails on the test passages or violates the tool policy \\
Sandbox & 300 s per program over the pool of one topic on DL 2019--2021 and 900--1,800 s on the held-out collections; 4 GB memory; no network, files or subprocesses \\
Pointwise judges & \texttt{claude-sonnet-5-5} with thinking type \texttt{between\_tools}; \texttt{claude-haiku-5-5} without thinking; effort medium; at most 200 output tokens; temperature not sent \\
Listwise judge & up to 100 passages per call, 15 output tokens per passage, up to 2 follow-up calls for passages without a grade \\
Seed & 20261017, for the pool order and the unlabeled passages in the writer prompt \\
Prices & USD per million input / output tokens: Opus 4 / 20, Sonnet 2 / 10, Haiku 0.10 / 0.50 \\
\bottomrule
\end{tabular}
\end{adjustbox}
\end{table}

\FloatBarrier
\begin{figure*}[p]
\lstset{basicstyle=\ttfamily\fontsize{4.6pt}{5.3pt}\selectfont}
\begin{minipage}[t]{0.495\textwidth}
\textbf{\footnotesize Writer prompt for TREC DL} {\footnotesize (\texttt{writer\_p1\_levels.md}). Placeholders are in double braces.}
\begin{lstlisting}
You will write a Python program that assigns relevance grades to passages for one search query. The program replaces a human relevance assessor for this query: it will be run once over every passage in the judging pool, with no further help from you.

Query:
{{query}}

{{narrative_block}}{{scale}}
{{calibration}}
{{labeled_examples}}

The program receives the whole judging pool for this query: {{pool_size}} passages, in random order. The pool was formed from the top results of many retrieval systems, so most passages share words with the query; grades should separate passages that answer the query from passages that only mention its topic. Grades may use statistics over the pool (for example a passage's BM25 percentile within the pool, or its term coverage relative to the best-covered passage).

Here are {{n_examples}} passages from this query's pool, unlabeled, to show the text format:
{{examples}}

{{tool_policy}}

Write exactly this function:

def assess(query: str, results: list[dict]) -> list[dict]:
    """results: list of {"id": str, "text": str}. Return one dict per passage, in the same order:
    {"on_subject": bool, "some_answer": bool, "exact_answer": bool, "dedicated": bool}"""

The grade is computed outside your code from these four answers, following the grade definitions: grade 0 when on_subject is false; grade 1 when on_subject is true and some_answer is false; grade 3 when some_answer, exact_answer and dedicated are all true; grade 2 in the other cases where some_answer is true.
- on_subject: the passage is about the specific subject of the query. Sharing words or the general topic with the query does not make a passage on_subject.
- some_answer: the passage contains at least part of the answer the query asks for.
- exact_answer: the passage states the exact answer.
- dedicated: the passage is mainly about the query; an answer hidden among unrelated material does not make a passage dedicated.

Set a value to true only when the passage gives explicit evidence for it. When the evidence is absent or uncertain, set it to false, so that the grade goes down. jk.detail and jk.unrelated_share measure how much of a passage supports the query and how much is unrelated.

Specific to this query, decide which entities, concepts, answer types (a number, date, definition, procedure, list) and wording establish each of the four answers, and encode those decisions in the program. Write constants such as synonym lists and answer patterns into the program. Do not define a function named label. The function must return a list of len(results) dicts and must not raise on empty or unusual text.

Reply with one ```python code block containing the complete program and nothing else.
\end{lstlisting}
\end{minipage}\hfill
\begin{minipage}[t]{0.475\textwidth}
\textbf{\footnotesize Filled-in blocks for DL 2019:} {\footnotesize \texttt{\{\{scale\}\}}, \texttt{\{\{calibration\}\}} and the start of \texttt{\{\{labeled\_examples\}\}}. \texttt{\{\{narrative\_block\}\}} is empty on DL.}
\begin{lstlisting}
Grades (TREC Deep Learning passage judgments):
3 = perfectly relevant: the passage is dedicated to the query and contains the exact answer.
2 = highly relevant: the passage has some answer for the query, but the answer may be a bit unclear, or hidden among extraneous information.
1 = related: the passage seems related to the query but does not answer it.
0 = irrelevant: the passage has nothing to do with the query.

Grade distribution: in past TREC Deep Learning judging pools, 64% grade 0, 17% grade 1, 13% grade 2, 6% grade 3. Most pooled passages share words with the query, and many passages that share words or the general topic with the query were graded 0. Use grade 1 only for passages that are about the specific subject of the query but do not answer it.

Examples graded by NIST assessors, from 3 other topics (one passage per grade; passages may be shortened):

Example topic 1. Query: why did the us volunterilay enter ww1
  Grade 3: The triggering cause was the continued sinking of neutral ships, which harmed American commercial interests. ...
  Grade 2: ...  Grade 1: ...  Grade 0: ...
\end{lstlisting}
\vspace{4pt}
\textbf{\footnotesize Paragraph added for \methodL{}} {\footnotesize after \texttt{\{\{labeled\_examples\}\}}. Each labeled passage is a line \texttt{[L$i$] NIST grade $g$:} followed by the passage text.}
\begin{lstlisting}
Here are {{n_labeled}} passages from this query's pool with the grades that NIST assessors gave them. They show where the assessors drew the lines between the grades for this query. The program will grade every passage in the pool, these included. Encode general rules for this query that give these passages their NIST grades; do not single out these passages by their wording.
{{topic_labeled}}
\end{lstlisting}
\end{minipage}
\caption{Writer prompt of the main method on TREC DL.}
\label{fig:prompt}

\vspace{2pt}
\lstset{basicstyle=\ttfamily\fontsize{4.6pt}{5.3pt}\selectfont}
\begin{lstlisting}
Tool policy. The program may import only:
- Python standard library modules: re, math, string, collections, itertools, functools, statistics, unicodedata, difflib, heapq, bisect, typing, dataclasses, operator, fractions, decimal, datetime, calendar, json, random (seed it if used).
- Lightweight libraries: numpy, scipy, rapidfuzz, rank_bm25, scikit-learn (linear models, SVM, naive Bayes, trees, gradient boosting), sklearn_crfsuite, dateparser, quantulum3, pint, snowballstemmer.
- The helper module `from pilot import judgekit as jk`, which provides:
  jk.tokenize(text) -> list[str]            lowercased alphanumeric tokens
  jk.terms(text, stemmed=True, drop_stopwords=True) -> list[str]
  jk.stem(token) -> str                      Snowball English stemmer
  jk.STOPWORDS                               frozenset of English stopwords
  jk.bm25_scores(query, texts, k1=0.9, b=0.4) -> list[float]   BM25 with statistics from `texts`
  jk.coverage(query, text) -> float          fraction of distinct query terms in text
  jk.idf_coverage(query, text, pool_texts) -> float
  jk.bigram_coverage(query, text) -> float
  jk.phrase_in(phrase, text) -> bool         exact phrase match on normalized tokens
  jk.fuzzy(a, b) -> float                    rapidfuzz partial_ratio in [0, 1]
  jk.numbers(text) -> list[float];  jk.years(text) -> list[int]
  jk.percentiles(scores) -> list[float]      rank of each score within the list, in [0, 1]
  jk.cut(values, cutoffs) -> list[int]       grade = number of increasing cutoffs at or below each value
  jk.nlp()                                   cached spaCy en_core_web_sm pipeline; jk.entities(text) -> [(text, label)]
  jk.sentences(text) -> list[str]           sentences split at . ! ? and line breaks
  jk.detail(terms, text, min_terms=2) -> int   number of sentences containing at least min_terms of the given
                                             words or phrases (stemmed); 2 or more means substance beyond a mention
  jk.unrelated_share(reference, text) -> float   share of the passage's words in sentences that share no content
                                             term with `reference` (the query or the narrative); 0 = all related
No neural models other than jk.nlp(), no LLMs, no embeddings, no network, no file access, no subprocesses,
no eval/exec/open. The program must be deterministic and must finish in under 60 seconds for 500 passages
(jk.nlp() costs about 1 ms per passage; use it only where it helps).
\end{lstlisting}
\caption{Tool policy (\texttt{\{\{tool\_policy\}\}} in Figure~\ref{fig:prompt}). It lists the allowed imports and the toolkit functions. The sandbox enforces the import rules.}
\label{fig:policy}
\end{figure*}

\FloatBarrier
\end{document}